\documentclass[%
 preprint,
nofootinbib,
 amsmath,amssymb,
 aps,
pre,
floatfix,
]{revtex4-1}
\usepackage{natbib,subfig}
\usepackage{graphicx}
\usepackage{dcolumn}
\usepackage{bm}
\usepackage{hyperref}
\begin{document}

\count\footins = 1000
\title{Origin of chaos near three-dimensional quantum vortices:\\ A  general Bohmian theory}
%
\author{Athanasios C. Tzemos}\email{thanasistzemos@gmail.com}
\author{Christos Efthymiopoulos}
\email{cefthim@academyofathens.gr}
\author{George Contopoulos}%
\email{gcontop@academyofathens.gr}
\affiliation{%
 Research Center for Astronomy and Applied Mathematics of the 
 Academy of Athens Soranou Efesiou 4 Athens, GR-11527
}%

\date{\today}

\begin{abstract}
We provide a general theory for the structure of the quantum flow near 
3-d `nodal lines', i.e. one-dimensional loci where the 3-d wavefunction 
becomes equal to zero. In suitably defined co-ordinates (co-moving with 
the nodal line) the generic structure of the flow implies the formation 
of 3-d quantum vortices. We show that such vortices are accompanied by 
nearby invariant lines of the co-moving quantum flow, called 'X-lines', 
which are normally hyperbolic. Furthermore, the stable and unstable 
manifolds of the X-lines produce chaotic scatterings of nearby quantum 
(Bohmian) trajectories, thus inducing an intricate form of the quantum 
current in the neighborhood of each 3-d quantum vortex. Generic formulas 
describing the structure around 3-d quantum vortices are provided, 
applicable to an arbitrary choice of 3-d wavefunction. We also give 
specific numerical examples, as well as a discussion of the physical 
consequences of chaos near 3-d quantum vortices. 
\end{abstract}
\maketitle


\section{Introduction}
\label{sec:intro}
The dynamics of quantum systems with vortices is of both theoretical 
and experimental interest. Quantum vortices are generated around singular 
points of the complex wavefunction $\Psi$ and have been found in different 
physical problems with two or more dimensions \cite{benseny,   hirschfelder1974quantum}, e.g. 
field theory \cite{dirac1931quantised}, Bose-Einstein condensates 
\cite{fetter2001vortices}, decoherence theory \cite{sanz2007quantum}, chemical reaction 
dynamics  \cite{babyuk2011study}, wave-packet interference \cite{chou2009hydrodynamic},
 quantum tunneling \cite{bell2013tunnel}, molecular dynamics simulations 
 \cite{curchod2011trajectory, garashchuk2015approximate},  electron microscopy 
\cite{rudinsky2017novel}, superconductors \cite{atzmon2012alternating, 
fazio2013charges} etc.

The trajectory-based, or `Bohmian', approach 
\cite{madelung1927quantentheorie, Bohm, BohmII,trahan2005quantum, 
sanz2012trajectory, sanz2013trajectory}  lends itself conveniently to the study of 
quantum vortices. In this approach, we consider trajectories which trace the
 quantum-mechanical currents according 
to the nonlinear (`pilot-wave') equations of motion. The trajectories are "guided" 
by the wave-function $\Psi$ which evolves in time according to the usual 
Schr\"{o}dinger's equation. The 'ontological' content of the Bohmian theory, 
as well as the 'reality' of Bohmian trajectories, has been a matter of intense 
debate over the years, in view also of its connection to experimental results
or the interpretations of the results of statistical experiments 
(see for example \cite{deotto1998bohmian, vaidman2005reality,durr2009bohmian,kocsis2011observing,
mahler2016experimental,gisin2018bohmian}). On the other hand, a quite
distinct motivation for studying the properties of Bohmian
trajectories stems from the standpoint of applications of the \textit{trajectory
methods} in the quantum dynamics of  particular physical systems
(see \cite{benseny} or \cite{sanz2012trajectory, sanz2013trajectory}). In fact,
trajectory-based methods are useful in a variety of applications. In particular, the behavior of the trajectories close to quantum vortices presents 
special interest,   since it has been demonstrated that trajectories 
scattered by quantum vortices exhibit complex or {\it chaotic} behavior.  
The connection of chaos with the existence of nodal points of the 
wavefunction has been established since long \cite{wu1994inverse, frisk1997properties, 
wu1999quantum, falsaperla2003motion, wisniacki2005motion, wisniacki2007vortex,
efthymiopoulos2007nodal, Efth2009, cesa2016chaotic}. Besides its general 
importance in all the above mentioned applications, chaos plays a special 
role in the accuracy of methods aiming to compute the Schr\"{o}dinger 
evolution through quantum trajectories, as, for example, in hydrodynamical 
solvers of the Schr\"{o}dinger equation \cite{lopreore1999quantum, babyuk2003hydrodynamic}, 
or in the conditional wave function approach to the quantum N-body problem 
\cite{alarcon2013computation, marian2016noise,colomes2017quantum}. Finally, as reviewed in 
\cite{efthymiopoulos2017chaos}, chaos may play a crucial role in a 
more theoretical framework, i.e., the question of how the rules of quantum 
probabilities (Born's rule) can be shown to emerge through the statistical
properties of quantum trajectories \cite{valentini2005dynamical, efthymiopoulos2006chaos} 
(see also \cite{durr1992quantum, durr2009bohmian}). 

Regarding, now, the mechanisms which lead to the emergence of chaotic Bohmian
trajectories in
the presence of quantum vortices, in 2005 Wisniacki and Pujals and Borondo et al. 
(see \cite{wisniacki2005motion, borondo2009dynamical}) were
the first to emphasize that it is the motion of quantum nodes which 
is responsible for the generation of chaos. In 
(\cite{wisniacki2005motion, borondo2009dynamical}) the emergence of chaos is described by the mechanism of homoclinic tangle.  
In \cite{Efth2009}, analytical formulas are provided showing that the emergence of a `nodal point-Xpoint'  topology close to any node is general, i.e. it applies for arbitrary choices of quantum model and wavefunction, while chaos is described as a `scattering' phenomenon. The connection of the scattering phenomenon with the homoclinic approach (\cite{wisniacki2005motion, borondo2009dynamical}) is an open problem.

The mechanisms of chaos discussed in  (\cite{wisniacki2005motion}, \cite{borondo2009dynamical}) and \cite{Efth2009}  refer to 2-d quantum systems. To our knowledge 
no such study of the mechanism of chaos has yet been presented in the case of three dimensional 
systems. This is, precisely, our scope in the present work.

In our work below we establish generic formulas which hold 
for the description of the quantum flow close to {\it nodal lines}, 
i.e. one dimensional curves representing continuous families of nodal 
points in the 3-d configuration space. These formulas justify 
theoretically results found in specific numerical examples in two 
previous papers of ours \cite{Tzemos2016, contopoulos2017partial}. 
The key remark, originally due to \cite{falsaperla2003motion}, is that 
in the neighborhood of a nodal line, the quantum flow is arranged in a 
foliation of surfaces which are nearly planar and normal to the nodal 
line. Following \cite{falsaperla2003motion}, these are called 
below the `Falsaperla-Fonte planes', or simply `F-planes'. Since 
sufficiently close to a nodal point the quantum flow becomes essentially 
planar (i.e. nearly confined in a F-plane), the generic picture found 
in \cite{Efth2009} for vortices in 2-d systems can be generalized in 
the 3-D case as well, with a number of additional considerations (or 
terms, in the formulas) accounting for the distortion of the quantum 
flow with respect to a perfectly planar form. 

\begin{figure}
\includegraphics[scale=0.4]{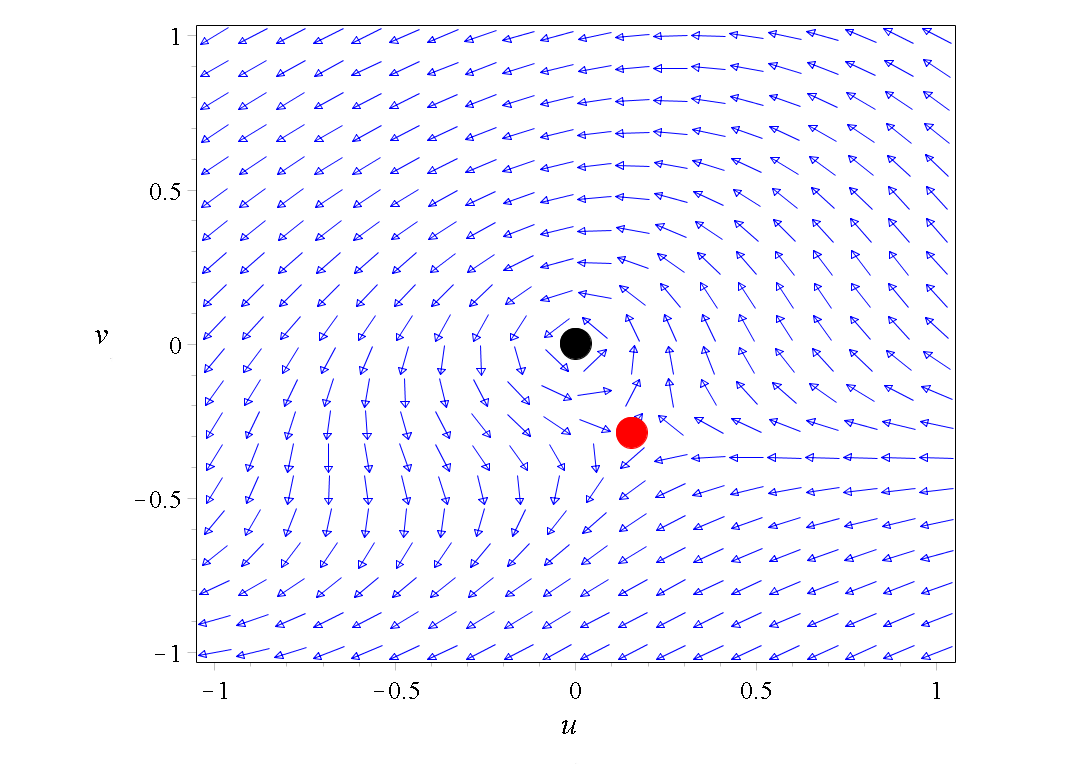}
\caption{The local quantum flow on the F-plane around a nodal point.
In coordinates co-moving with the nodal point,  the flow very 
close to the nodal point (black dot) has the shape of a spiral, while 
further away there is an X-point (red dot) which has zero velocity 
with respect to the nodal point.}
\label{fig:npxpc}
\end{figure}

According to this picture, the projection of the quantum flow in the 
F-plane around every nodal point along one nodal line takes the form 
shown schematically in Fig.\ref{fig:npxpc}. This figure illustrates 
the quantum flow in the F-plane as viewed in a frame of reference 
{\it co-moving with a nodal point}. We find that every nodal 
point is necessarily accompanied by a single second critical point of 
the comoving flow, called 'X-point', whose linear stability character 
is hyperbolic. In the rest frame, this is a point in space where the 
quantum trajectory has instantaneous velocity equal to that of the 
nodal point. The whole structure, as viewed in the co-moving frame, is 
called a `nodal point-X-point complex' (NPXPC). Taking the foliation 
of all NPXPCs along a nodal line forms a `3-d structure of NPXPCs'. 
 This structure complements in a substantial 
way the `cylindrical structure' observed in \cite{Tzemos2016}, 
by adding a continuous family of X-points to it, which we call the 
`X-line'. This is a one-dimensional critical curve of the 
flow as viewed in a frame co-moving with the nodal line (see precise 
definitions in section \ref{sec:genthe} and Appendix I). As such, 
the `X-line' constitutes a normally hyperbolic manifold 
\cite{wiggins2013normally}, wherefrom 2-dimensional unstable and 
stable manifolds emanate.  As shown below, one branch of these 
manifolds terminates in the nodal  line through spirals, while the remaining 
branches asymptotically deviate away from the X-line. 

The so-formed 3-d structure gives a complete characterization of the 
3-d quantum vortex around a nodal line. More importantly, 
trajectories which come close to the X-line undergo hyperbolic 
scattering with features similar to those described in \cite{Efth2009} 
for the 2-D case. As a result, the local value of the Lyapunov exponent 
(or `stretching number' \cite{voglis1994invariant, Contopoulos200210}) 
undergoes positive 
jumps in every such scattering, hence accumulating to a positive 
Lyapunov characteristic exponent, i.e., chaos. This latter result 
is verified by numerical experiments, 
as described in section \ref{sec:num}, hence validating the theory
of section \ref{sec:genthe} below.  

The paper is structured as follows: in section \ref{sec:genthe} we develop 
the general theory of the structure of the quantum flow (and trajectories) 
near quantum vortices, supplemented with a generic set of formulas which 
apply for arbitrary 3-d wavefunctions possessing nodal lines. In section 
\ref{sec:num} we validate the theoretical analysis of the previous section 
by numerical experiments, computing a real (non-schematic) example of 
`the 3-d structure of NPXPCs', and probing the correlation between 
jumps in the local Lyapunov exponents of trajectories and close 
encounters with the X-line. Section \ref{sec:concl} summarizes the 
basic conclusions of the present study.

\section{3-d structure of nodal and X points: general theory}
\label{sec:genthe}
\subsection{Summary of results}
\label{subsec:summary}
Consider a 3-d quantum system with wavefunction $\Psi(x,y,z,t)$, 
at the time $t$, where $(x,y,z)$ are Cartesian space co-ordinates. 
The quantum current is given by $\mathbf{j}=\frac{\hbar}{2mi}(\Psi^*\nabla\Psi
-\Psi\nabla\Psi^*)$ ($\hbar=$Planck's constant, $m=$ the particle mass, 
$\hbar=m=1$ in the sequel). The quantum current can be traced by a swarm 
of `Bohmian' trajectories defined via the flow $\mathbf{v}=\mathbf{j}/|\Psi|^2$. 
These equations of motion take the generic form
\begin{equation}\label{eq:bohmgen}
\frac{dx}{dt}=f_x(x,y,z,t),\,\,\,\frac{dy}{dt}=f_y(x,y,z,t),\,\,\,\frac{dz}{dt}=f_z(x,y,z,t)~~.
\end{equation}
Eqs.(\ref{eq:bohmgen}) constitute a non-autonomous 3-d dynamical system. 
Numerical investigations in various systems of the form (\ref{eq:bohmgen})
have revealed the typical existence of chaotic trajectories (see, for 
example \cite{frisk1997properties, falsaperla2003motion, Tzemos2016}. 
Our purpose is to explain the dynamical origin of the 
trajectories' chaotic behavior. To this end, it proves crucial to focus
on the structure of the flow induced by Eqs.(\ref{eq:bohmgen}) near 
loci in space where $\Psi(x,y,z,t)=0$. These loci are the `nodal lines' 
of the wavefunction $\Psi$. 

We invoke the following two approximations, further specified in 
subsequent sections: 

i) Under an appropriate `adiabatic' condition (see subsection 
\ref{subsec:npxpc}), the flow (\ref{eq:bohmgen}) in a neighborhood of 
every point along a nodal line, and within a given time interval, can be 
approximated as nearly autonomous   in a set of variables $(u,v,w)$, 
found after a linear transformation ${\cal L}: (x,y,z)\rightarrow(u,v,w)$. 
The variables $(u,v,w)$ correspond physically to co-ordinates locally 
attached to and co-moving with each separate nodal point along a nodal line. 

ii) One can use the above co-ordinates in order to construct a new 
dynamical system of the form:
\begin{equation}\label{eq:eqmouvs}
\frac{dU}{d\tau} = F_U(U,V,S),~~ 
\frac{dV}{d\tau} = F_V(U,V,S),~~ 
\frac{dS}{d\tau} = F_S(U,V,S) 
\end{equation}
where the variables $(U,V,S)$ are provided by a transformation 
\begin{equation}\label{eq:uvstra}
(U,V,S)=(u,v,w)+{\cal NL}(u,v,w)
\end{equation}
${\cal NL}(u,v,w)$ being a small nonlinear correction to the identity. 
The new time $\tau$ corresponds to a suitably defined re-parametrization 
of the original time $t$. 

Using the above approximations, we deduce the following, which is the 
main result of the present paper: We show that the equations of motion 
(\ref{eq:eqmouvs}) admit as solution an invariant continuous family of 
hyperbolic points, which, together, form a normally hyperbolic `X-line'. 
Furthermore, the stable and unstable manifolds of the X-line connect 
with the associated nodal line through a family of spirals, forming 
a 3-d vortex. This we call the `3-d structure of nodal point - X-point 
complexes'. The geometric structure of these manifolds serves to explain 
the hyperbolic scattering of nearby trajectories, i.e., the appearance of 
chaos near 3-d quantum vortices. In particular, it is the dynamics close to 
the X-line, rather than close to the nodal line, which is responsible for chaos. 
This latter fact is substantiated with numerical experiments in section 
\ref{sec:num}. 

The complete construction of the transformation (\ref{eq:uvstra}), as well 
as the Eqs.(\ref{eq:eqmouvs}), is given in Appendix I. Most features of 
these equations can be approximated by a local construction, which yields  
equations of motion around only one nodal point along a nodal line. We now 
focus on the latter construction, which is similar as the one introduced   
in \cite{Efth2009} for 2-d systems. 

\subsection{Nodal line}
\label{subsection:nodline}
Let $\Psi_R(x,y,z,t)=\Re(\Psi(x,y,z,t))$, $\Psi_I(x,y,z,t)
=\Im(\Psi(x,y,z,t))$. We assume that, at fixed time $t$, the $2\times 3$ 
set of equations
\begin{equation}\label{eq:nodal}
\Psi_R(x,y,z,t)=0,~~~\Psi_I(x,y,z,t)=0
\end{equation}
admits solutions lying in one or more curves in the space $(x,y,z)$. 
We call such curves the `nodal lines' of the wavefunction $\Psi$ at 
the time $t$. 

Depending on the wavefunction under study, the shape of nodal lines at a 
given time $t$ can be quite complex, as it can be composed of different 
branches, possibly forming loops or extending to infinity. On the other hand, it is possible to introduce a continuous 
parametrization of one nodal line in every open segment of it. To this 
end, let O be an arbitrary point along a nodal line not belonging to its 
boundary (we call O the `origin'). In an open neighborhood ${\cal V}_0$ 
of $O$, we define $s$ to be the length parameter along the curve, starting 
from O and with sign indicating the direction of inscription of the line 
with respect to O. For every point of the nodal line $(x_0,y_0,z_0)$ 
belonging to ${\cal V}_0$ we can then assign a unique value of $s$, 
thus defining the continuous and smooth functions $x_0(s),y_0(s),z_0(s)$.   

Setting $\Psi = \Psi_R + i\Psi_I$, the Bohmian equations of motion 
in the inertial frame read (with $m=\hbar=1$):
\begin{equation}\label{eq:bohm}
\mathbf{v}=\dot{\mathbf{r}}=\frac{\Psi_R\nabla\Psi_I - 
\Psi_I\nabla\Psi_R}{\Psi_R^2+\Psi_I^2}.
\end{equation}
Since $\Psi_R(x_0(s),y_0(s),z_0(s),t) 
= \Psi_I(x_0(s),y_0(s),z_0(s),t)=0$ 
one has
\begin{equation}\label{eq:gradpsi}
\frac{d\Psi_R}{ds}=\nabla\Psi_R\cdot\frac{d\mathbf{r_0}}{ds}=0, \,\,\,\,
\frac{d\Psi_I}{ ds}=\nabla\Psi_I\cdot\frac{d\mathbf{r_0}}{ds}=0 
\end{equation}
and, hence, $\nabla\Psi_R\perp d\mathbf{r_0}/ds$, 
$\nabla\Psi_I\perp d\mathbf{r_0}/ds$ at all points 
$\mathbf{r_0}\equiv(x_0,y_0,z_0)$. Then, by Eq.(\ref{eq:bohm}) we deduce 
that the Bohmian velocity field tends to become normal to the nodal curve 
as we tend towards any point on this curve. The plane normal to a nodal 
curve at one of its nodal points is  called the `F-plane' 
associated with the nodal point (see \cite{falsaperla2003motion}). 

Assuming the wavefunction to be continuous and with continuous 
derivatives in ${\cal V}_0$ at all times between $t$ and $t+\Delta t$, 
where $\Delta t$ is a small time interval, Eqs.(\ref{eq:nodal}) imply 
that the nodal line evolves smoothly in time within ${\cal V}_0$. 
In particular, let ${\cal F}_{\mathbf{r}_0,t}$ be the F-plane associated 
with the nodal point $\mathbf{r}_0$ at the time $t$, and $\mathbf{r}_0(t')$ 
be the point where the nodal line at a later time $t'$ intersects the same 
F-plane ${\cal F}_{\mathbf{r}_0,t}$, for all $t'$ with 
$t\leq t'\leq t+\Delta t$. We define the velocity $\mathbf{V}_0$ of the 
nodal point $\mathbf{r}_0$ at the time $t$ as 
\begin{equation}\label{eq:nodalvel}
\mathbf{V}_0(t) = \lim_{t' \rightarrow t}
\frac{\mathbf{r}_0(t')-\mathbf{r}_0(t)}{ t'-t}~~.
\end{equation}
Clearly, the velocity vector $\mathbf{V}_0(t)$ belongs to 
${\cal F}_{\mathbf{r}_0,t}$, thus it is also perpendicular to the nodal 
line. 

We note here that, as it will be discussed below, `X-lines' necessarily appear only 
when we consider the quantum flow around {\it time-varying} nodal lines. 
Similarly to the 2-d case (see \cite{efthymiopoulos2007nodal,Efth2009}),
in order to describe this phenomenon we must consider the form of the quantum
flow in a co-ordinate frame co-moving with the nodal line. In turn, this 
requires assigning a value of the velocity to every point along the nodal 
line. However, contrary to the 2-d case, in the 3-d case such assignment 
cannot be done unambiguously on the sole basis of the shape evolution of
the nodal line at nearby times, since there are infinitely many ways 
to continuously map each point of the nodal line, at time $t$, to a 
point on the deformation of the same nodal line at a later time $t'>t$. 
Using the intersection of the nodal line with the F-plane 
${\cal F}_{\mathbf{r}_0,t}$ allows to bypass this ambiguity. 
\footnote{One may also remark that the Bohmian equations (\ref{eq:bohm})  
cannot be used directly to define $\mathbf{V}_0(t)$, since the r.h.s. of 
these equations are singular at $\mathbf{r}=\mathbf{r}_0$ at the time $t$.}

\subsection{Structure of the flow near nodal points: quantum vortices}
\label{subsec:vortices}
\begin{figure}
\centering
\includegraphics[scale=0.4]{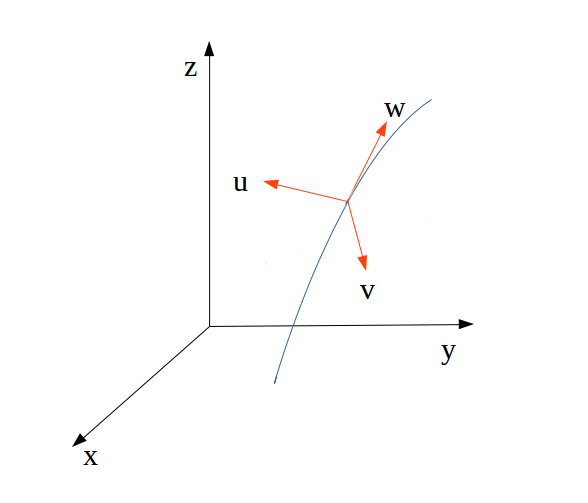}
\caption{Inertial (black) and  moving  (red) frames of reference.
 The axis $\mathbf{w}$ is tangent 
locally to the nodal line (blue curve), while the axes $\mathbf{u,v}$ 
lie in the F-plane.}\label{frofref}
\end{figure}

Let $\mathbf{r}_0=(x_0,y_0,z_0)$ be a particular nodal point at the time 
$t=t_0$. We now consider a coordinate change $(x,y,z)\rightarrow(u,v,w)$ 
such that: i) $(x_0,y_0,z_0)\rightarrow(0,0,0)$, i.e. the node is at the 
origin, ii) the new  co-ordinates are such that the w-axis is tangent to 
the nodal line at the point $(x_0,y_0,z_0)$, while the $u$ and $v$ are 
directions in the associated F-plane defined via a pre-selected geometric 
rule. For example, if $\mathbf{t}$, $\mathbf{n}$, $\mathbf{b}$ are the 
tangent, normal and bi-normal unitary vectors of the nodal line at 
$\mathbf{r}_0$, for an arbitrary point $\mathbf{r}\equiv(x,y,z)$ we 
define the transformation (Fig.~\ref{frofref}):
\begin{equation}\label{eq:frenet}
u=\mathbf{\Delta r}\cdot\mathbf{n},~~
v=\mathbf{\Delta r}\cdot\mathbf{b},~~
w=\mathbf{\Delta r}\cdot\mathbf{t}
\end{equation}
where $\mathbf{\Delta r}=\mathbf{r}-\mathbf{r}_0$. Since Eqs.(\ref{eq:frenet}) 
become singular at points $\mathbf{r}_0$ where the curvature of the nodal 
line becomes zero, we can use, alternatively, the transformation:
\begin{equation}\label{eq:cart}
\begin{pmatrix}
u\\v\\w
\end{pmatrix}=
\begin{bmatrix}
    \sin{\phi_m} & -\cos{\phi_m} & 0   \\
    \cos{\theta_m}\cos{\phi_m} & \cos{\theta_m}\sin{\phi_m} & -\sin{\theta_m} \\
    \sin{\theta_m}\cos{\phi_m} & \sin{\theta_m}\sin{\phi_m} & \cos{\theta_m}
\end{bmatrix} 
\begin{pmatrix}
x-x_0\\
y-y_0\\
z-z_0
\end{pmatrix}
\end{equation}
with
\begin{align}
&\phi_m=\arctan(t_y/t_x)\\&
\theta_m=\arccos(t_z/|t|)
\end{align}
where $\mathbf{t}=(t_x, t_y, t_z)$. Equations (\ref{eq:cart}) become 
singular whenever $\mathbf{t}$ becomes parallel to the z-axis. In numerical 
simulations we use both (\ref{eq:frenet}) and (\ref{eq:cart}), ensuring 
to properly deal with the corresponding singularities. 

Expressing $\Psi$ in the new co-ordinates $\Psi\equiv\Psi(u,v,w,t)$, 
Eqs.(\ref{eq:nodal}) and (\ref{eq:gradpsi}) take the form 
$$
\Psi(0,0,0,t=t_0)=0,~~~
\frac{\partial\Psi}{\partial w}\Bigg|_{u=0,v=0,w=0,t=t_0}=0~~
$$

Then, Taylor-expanding $\Psi$ around the origin (the nodal point) and up to terms 
of second degree in the variables $u,v,w$ yields
\begin{eqnarray}\label{eq:psitay}
\Psi &= 
&a_{000}+(a_{100}+ib_{100})u+
(a_{010}+ib_{010})v+
(a_{200}+ib_{200})u^2+
(a_{020}+ib_{020})v^2\nonumber\\
&+&
(a_{002}+ib_{002})w^2+
(a_{110}+ib_{110})uv+
(a_{101}+ib_{101})uw+
(a_{001}+ib_{001})vw +\ldots
\end{eqnarray}
with real {\it time-dependent} coefficients $a_{ijk}(t)$, $b_{ijk}(t)$, 
$i,j,k=0,1,\dots$. One has $a_{000}(t_0)=0$. As in the corresponding 
analysis in the 2-d case (see \cite{Efth2009}), we now consider a frame of 
reference centered at the nodal point 
$\mathbf{r}_0$ and moving with uniform velocity $(V_u,V_v,V_w)$ 
equal to the instantaneous velocity $\mathbf{V}_0$ of the nodal point  
defined in Eq.(\ref{eq:nodalvel}). Setting $G=\Psi_R^2+\Psi_I^2
=(a_{100}u+a_{010}v)^2+(b_{100}u+b_{010}v)^2+\dots$, the equations of 
motion in the above frame read
\begin{eqnarray}\label{eq:bohm3}
\frac{du}{dt}&=&\frac{1}{G}\bigg[\Big(\frac{\partial \Psi_I}
{\partial u}\Psi_R-\frac{\partial \Psi_R}{\partial u}\Psi_I\Big)
-V_uG\bigg]\nonumber\\
\frac{dv}{dt}&=&\frac{1}{G}\bigg[\Big(\frac{\partial \Psi_I}
{\partial v}\Psi_R-\frac{\partial \Psi_R}{\partial v}\Psi_I\Big)
-V_vG\bigg]\\
\frac{dw}{dt}&=&\frac{1}{G}\bigg[\Big(\frac{\partial \Psi_I}
{\partial w}\Psi_R-\frac{\partial \Psi_R}{\partial w}\Psi_I\Big)
-V_wG\bigg]~~. \nonumber
\end{eqnarray} 
We have $V_w=0$, since, by definition, the velocity $\mathbf{V}_0$ 
is orthogonal to the nodal line. We also note that the r.h.s. of 
Eqs.(\ref{eq:bohm3}) depends explicitly on time through the time 
dependence of the wavefunction $\Psi$.

Due to Eq.(\ref{eq:psitay}), we readily find that, as $\Psi\to0$ 
at $t=t_0$, the numerators in the r.h.s. of the first two of the 
Eqs.(\ref{eq:bohm3}) tend to  zero by terms linear in $u,v$, while 
the denominator (equal to $G$) tends to zero by terms quadratic 
in $u,v$. Thus, at $t=t_0$, the velocities $du/dt$, $dv/dt$ in 
Eqs.(\ref{eq:bohm3}) tend to infinity as we tend towards the nodal 
point $\mathbf{r}_0$. This implies that one can always find a 
sufficiently small neighborhood of the nodal point in which the 
Bohmian trajectories are inscribed with  velocities much larger 
than $\mathbf{V}_0$. As in (\cite{Efth2009}) we then invoke an 
{\it adiabatic approximation} to approximate Eqs.(\ref{eq:bohm3}) 
as nearly autonomous, i.e., to `freeze' the time in the r.h.s. of 
Eqs.(\ref{eq:bohm3}) and replace the dynamical system (\ref{eq:bohm3}) 
with
\begin{eqnarray}\label{eq:bohm3aut}
\frac{du}{dt}&=&\left(\frac{1}{G}\bigg[\Big(\frac{\partial \Psi_I}
{\partial u}\Psi_R-\frac{\partial \Psi_R}{\partial u}\Psi_I\Big)
-V_uG\bigg]\right)_{t=t_0}\nonumber\\
\frac{dv}{dt}&=&\left(\frac{1}{G}\bigg[\Big(\frac{\partial \Psi_I}
{\partial v}\Psi_R-\frac{\partial \Psi_R}{\partial v}\Psi_I\Big)
-V_vG\bigg]\right)_{t=t_0} \\
\nonumber\frac{dw}{dt}&=&\left(\frac{1}{G}\bigg[\Big(\frac{\partial \Psi_I}
{\partial w}\Psi_R-\frac{\partial \Psi_R}{\partial w}\Psi_I\Big)-
V_wG\bigg]\right)_{t=t_0}\nonumber
\end{eqnarray}  
The exact conditions of validity of the above adiabatic approximation 
are similar to those established in \cite{Efth2009} (see also 
subsection \ref{subsec:npxpc}).  

Under the approximation (\ref{eq:bohm3aut}), the quantum flow in the 
moving frame of reference becomes singular for trajectories very close 
to the nodal point. However, such singularity is regularizable via a 
time transformation. Let $(u(t),v(t),w(t))$ be a trajectory of 
Eqs.(\ref{eq:bohm3}). Let $t\rightarrow\tau$ be the time transformation 
defined via the differential equation $d\tau/dt=G(u(t),v(t),w(t))$. 
In the new time $\tau$, the trajectories of the  system 
(\ref{eq:bohm3}) are given by the `reduced' dynamical system
\begin{eqnarray}\label{eq:bohm3b}
\frac{du}{d\tau}&=&\Big(\frac{\partial \Psi_I}{\partial u}\Psi_R
-\frac{\partial \Psi_R}{\partial u}\Psi_I\Big)
-V_uG\label{bohm1b}\nonumber\\
\frac{dv}{d\tau}&=&\Big(\frac{\partial \Psi_I}{\partial v}\Psi_R
-\frac{\partial \Psi_R}{\partial v}\Psi_I\Big)
-V_vG\label{bohm2b}\\
\frac{dw}{d\tau}&=&\Big(\frac{\partial \Psi_I}{\partial w}\Psi_R
-\frac{\partial \Psi_R}{\partial w}\Psi_I\Big).\nonumber
\end{eqnarray}
 We note, in particular, that the system 
(\ref{eq:bohm3b}) shares the same critical points and geometrical 
trajectories with the system (\ref{eq:bohm3}), as can be deduced, 
e.g., by dividing each of the first two equations in (\ref{eq:bohm3}) 
and (\ref{eq:bohm3b}) by the third equation respectively. Hereafter we 
work with the reduced equivalent dynamical system (\ref{eq:bohm3b})
which is devoid of infinities in the close neighbourhood of 
the nodal point. This improves  significantly  the precision of 
numerical calculations of trajectories, where we can implement time 
regularization in the integration scheme.  Also, by freezing the new time
$\tau=\tau_0$ in the r.h.s. of Eqs.(\ref{eq:bohm3b}), one obtains the 
same approximate trajectories as in the system (\ref{eq:bohm3aut}).

Returning to our analysis, the nodal point 
$\mathbf{r}_0$, i.e., $u_0=v_0=w_0=0$ becomes a stationary point of the 
quantum flow as represented in the  `frozen time' approximation of Eqs.(\ref{eq:bohm3b}). 
In the second order approximation, these equations (\ref{eq:bohm3b}) read:
\begin{eqnarray}\label{eq:bohmexp}
\frac{du}{d\tau}&=& -Av+A_{200}{u}^{2}+ A_{020}{v}^{2}+
 A_{002} {w}^{2}+A_{110} uv+A_{011} vw+\dots\nonumber\\
\frac{dv}{d\tau}&= &Au+  B_{200} {u}^{2}+B_{020} {v}^{2}
+B_{002}{w}^{2}+B_{101} uw+ B_{110} uv+\dots\\
\frac{dw}{d\tau}&=& C_{200} {u}^{2}+C_{020} {v}^{2
}+ C_{110}uv+ C_{101} uw+C_{011}vw+\dots~~,\nonumber
\end{eqnarray}
with 
$$
A=\left( a_{100}b_{010}-a_{010}b_{100} \right)
$$
$$
A_{011}=\left( a_{010}b_{101}+a_{011}b_{100}-a_{100}b_{011}
-a_{101}b_{010} \right)
$$
$$
A_{110}=2\left( a_{010}b_{200}-a_{200}b_{010}\right)
-2V_u\left(a_{100}a_{010}+b_{100}b_{010}\right)
$$
$$
A_{020}=a_{010}b_{110}+a_{020}b_{{100}}
-a_{100}b_{020}-a_{110}b_{010}
-V_u\left(a_{010}^2+b_{010}^2\right)
$$
$$
A_{002}=\left( a_{002}b_{100}-a_{100}b_{002} \right)
$$
$$
A_{200}=a_{100}b_{200}-a_{200}b_{100}
-V_u\left(a_{100}^2+b_{100}^2\right)
$$
$$
B_{200}=a_{100}b_{110}-a_{110}b_{100}
+a_{200}b_{010}-a_{010}b_{200}
-V_v\left(a_{100}^2+b_{100}^2\right)
$$
$$
B_{002}= \left( a_{002}b_{010}-a_{010}b_{002} \right)
$$
$$B_{020}=a_{010}b_{020}-a_{020}b_{010}
-V_v\left(a_{010}^2+b_{010}^2\right)
$$
$$
B_{101}=\left(a_{100}b_{011}-a_{011}b_{100}+a_{101}b_{010}
-a_{010}b_{101} \right)=-A_{011}
$$
$$
B_{110}=2\left(a_{100}b_{020} -a_{020}b_{100}\right)
-2V_v\left(a_{100}a_{010}+b_{100}b_{010}\right)$$
$$
C_{200}=\left( a_{100}b_{101}-a_{101}b_{100} \right)
$$
$$
C_{020}= \left( a_{010}b_{011}-a_{011}b_{010} \right)
$$
$$
C_{011}=2\left( a_{010}b_{002} -a_{002}b_{010}\right)=-2B_{002}
$$
$$
C_{101}=2\left( a_{100}b_{002}-a_{002}b_{100}\right)=-2A_{002}
$$
$$
C_{110}=\left( a_{010}b_{101}-a_{101}b_{010}
+a_{100}b_{011}-a_{011}b_{100} \right)
$$
where $a_{ijk}, b_{ijk}$ are the coefficients of the wavefunction 
expansion (\ref{eq:psitay}) at the time $t=t_0$. We observe the basic 
symmetries $A_{011}=-B_{101}, C_{101}=-2A_{002}, C_{011}=-2B_{002}$. 
\footnote{We note that in quantum systems described by wavefunctions 
of the form:
$$
\Psi(\mathbf{x},t)=e^{\sigma(\mathbf{x})}\phi(\mathbf{x},t)
$$
where $\sigma(\mathbf{x})$ a real-valued function, a more precise 
expansion of the equations of motion can be found: 
$$
\dot{\mathbf{x}}
=\Im\Big(\frac{\nabla\Psi(\mathbf{x},t)}{\Psi(\mathbf{x},t)}\Big)
=\frac{\nabla\Psi_I(x,t)\Psi_R(\mathbf{x},t)
-\nabla\Psi_R(x,t)\Psi_I(x,t)}{\Psi_R^2(x,t)+\Psi^2_I(\mathbf{x},t)}
$$
$$
=\frac{\nabla\phi_I(\mathbf{x},t)\phi_R(\mathbf{x},t)
-\nabla\phi_R(\mathbf{x},t)\phi_I(\mathbf{x},t)}{\phi_R^2(\mathbf{x},t)
+\phi^2_I(\mathbf{x},t)}
$$
where $\phi(\mathbf{x},t)=\phi_R(\mathbf{x},t)+i\phi_I(\mathbf{x},t)$. 
In such systems, one obtains the same formulas as in 
Eq.(\ref{eq:bohmexp}), where the coefficients $a_{ijk},b_{ijk}$ refer 
to the Taylor expansion, as in  Eq.(\ref{eq:psitay}), but of the 
function $\phi$ instead of $\Psi$. All symmetries of the coefficients 
and consequent results apply equivalently for the functions $\phi$ 
and $\Psi$. In numerical computations, however,using the coefficients 
of $\phi$ yields results more precise than those found with the 
coefficients of $\Psi$.}

Linearization of the system (\ref{eq:bohm3b}) around the nodal 
point yields the equations
\begin{equation}\label{eq:nodlin}
\frac{du}{d\tau}= -Av,\quad
\frac{dv}{d\tau}= Au,\quad
\frac{dw}{d\tau}= 0~~.
\end{equation}
The eigenvalues of the associated Jacobian matrix are $(iA,-iA,0)$. 
Thus, to the lowest order approximation, the nodal point is a nilpotent 
center of the flow. The eigenvector associated with the zero eigenvalue 
is $(0,0,1)$, which corresponds to a vector tangent to the nodal line 
at the nodal point $\mathbf{r}_0$, while the center manifold of the 
system (\ref{eq:nodlin}) coincides with the F-plane ${\cal F}_0$. 

In the linear approximation, the trajectories in ${\cal F}_0$ are circles 
described around the nodal point with frequency $\omega = A$. However, 
due to the fact that all eigenvalues have null real part, these linear 
solutions are not structurally stable with respect to nonlinear 
perturbations (see \cite{strogatz2014nonlinear}). Thus, to determine the character of the 
trajectories we must consider higher than linear order terms of the system 
(\ref{eq:bohmexp}). 

As in \cite{Efth2009} we transform Eqs.(\ref{eq:bohmexp}) to cylindrical 
coordinates $R,\phi,w$, by setting $u=R\cos\phi, v=R\sin\phi$. This leads 
to a system of three ordinary differential equations which contain 
coupling terms polynomial in $R,w$ and trigonometric polynomial in the 
azimuth $\phi$. Explicit solutions within a a bounded domain around the 
nodal point $\mathbf{r}_0$ can be constructed under the following 
approximations: assume the wavefunction $\Psi$ has support over a volume 
in configuration space of linear size ${\cal L}$. Consider a 
sphere ${\cal S}_\epsilon$ of radius $\epsilon$ centered around the 
nodal point $\mathbf{r}_0$, where $\epsilon$ is chosen such that, 
for all initial conditions within ${\cal S}_\epsilon$, the second-order 
truncation of the equations of motion (Eqs.(\ref{eq:bohmexp})) yields 
solutions  differing from the solutions of the complete system by less 
than a prescribed accuracy up to a maximum time of interest. Let now $R_0$ 
be the radius of curvature of the nodal line at the point $\mathbf{r}_0$. 
Consider the open domain ${\cal V}_\epsilon$ defined by all points within 
${\cal S}_\epsilon$ such that ${\cal L}|w|<\epsilon^{3/2}R_0^{1/2}$. 
It can be shown that, for all points within the domain ${\cal V}_\epsilon$, 
there are positive constants $C_1={\cal O}({\cal L}^{-9/2})$, $C_2
={\cal O}({\cal L}^{-7/2})$ such that the following relations hold
\begin{equation}\label{eq:curve}
\max\left(|a_{002}|,|b_{002}|\right)|w|^2<C_1\epsilon^3,~~~
\max\left(|a_{002}|,|b_{002}|\right)|w|<C_2\epsilon^{3/2}R_0^{-1/2}~,
\end{equation}
where $a_{002}$ and $b_{002}$ are the coefficients of the wavefunction 
expansion appearing in (\ref{eq:psitay}). 
\footnote{
Assuming $\mathbf{r}_0$ not to be an inflection point of the nodal line, 
the nodal line close to $\mathbf{r}_0$ has the form of a parabola $\gamma$, 
which can be represented by a parametrization in terms of functions 
$u_\gamma(w), v_\gamma(w)$ as:
$$
u_\gamma(w) =\frac{1}{2}g_u w^2 +{\cal O}(w^3),~~~
v_\gamma(w) =\frac{1}{2}g_v w^2 +{\cal O}(w^3),~~~
$$
with $g_u=d^2u_\gamma/dw^2|_{w=0}$, $g_v=d^2v_\gamma/dw^2|_{w=0}$. 
One has $g_u={\cal O}\left(R_0^{-1}\right)$, 
$g_v={\cal O}\left(R_0^{-1}\right)$. Using the above equation, as well 
as the Taylor expansion (\ref{eq:psitay}), Eq.(\ref{eq:gradpsi}) 
yields, to the lowest order (linear in $w$), the set of equations:
$$
2a_{002}+a_{100}g_u+a_{010}g_v=0,~~~ 
2b_{002}+b_{100}g_u+b_{010}g_v=0~~~. 
$$
Restoring units ($\Psi\sim {\cal L}^{-3/2}$), we find 
$a_{002}={\cal O}\left({\cal L}^{-7/2}/R_0\right)$, 
$b_{002}={\cal O}\left({\cal L}^{-7/2}/R_0\right)$, 
which leads to the inequalities (\ref{eq:curve}). }
In normalized units (${\cal L}\sim 1$), the first of Eqs.(\ref{eq:curve}) 
implies that, within the domain ${\cal V}_\epsilon$, every term containing 
a factor $A_{002}w^2$ or $B_{002}w^2$ in the first two of 
Eqs.(\ref{eq:bohmexp}) is of order ${\cal O}(\epsilon^3)$, while the 
remaining terms are of order ${\cal O}(\epsilon^2)$. Ignoring the terms 
$A_{002}w^2$ or $B_{002}w^2$, and transforming to cylindrical co-ordinates, 
the first two of Eqs.(\ref{eq:bohmexp}) then take the form
\begin{eqnarray}\label{eq:cylrphi}
\frac{dR}{d\tau}&=&\frac{(u\dot{u}+v\dot{v})}{R}=
c_2(\phi)R^2+\dots\nonumber\\
\frac{d\phi}{d\tau}&=&\frac{(u\dot{v}-v\dot{u})}{R^2}=
d_0+d_1(\phi)R+h_1w+\dots
\end{eqnarray}
where the coefficients $c_2(\phi)$, $d_1(\phi)$ are odd trigonometric 
functions of $\phi$, $d_0=A=const$, and $h_1=B_{101}$. The last relation 
is a consequence of the symmetry $A_{011}=-B_{101}$, and plays a crucial 
role below. Finally, the third of Eqs.(\ref{eq:bohmexp}) takes the form
\begin{equation}\label{eq:cylw}
\frac{dw}{d\tau}=e_2(\phi)R^2+k_2(\phi)R w+\dots
\end{equation}
where $e_2(\phi)$ and $k_2(\phi)$ are  trigonometric functions of $\phi$, 
even and odd respectively. 

Equations (\ref{eq:cylrphi}) and (\ref{eq:cylw}) still contain nonlinear 
couplings between the co-ordinates $(R,\phi,w)$, thus obstructing obtention 
of an explicit solution. As in \cite{Efth2009}, we then compute an `averaged' 
system of equations which possesses an explicit solution. Dividing the 
first with the second of Eqs.(\ref{eq:cylrphi}), and averaging over the 
angle $\phi$ we find the average dependence of $R$ on $\phi$ (denoted 
$\overline{R}$) along the quantum flow in the neighborhood of 
$\mathbf{r}_0$ as:
$$
\frac{d\overline{R}}{d\phi} = \frac{1}{ 2\pi}\int_0^{2\pi}
\left(\frac{c_2(\phi)R^2+\ldots}{
d_0+d_1(\phi)R+h_1w+\dots}\right)d\phi~~.
$$
Expanding the denominator in the quantities $d_1/d_0$, $h_1/d_0$ and 
taking into account the parity and form of the various trigonometric 
coefficients defined above leads to
\begin{equation}\label{eq:drdphif3}
\frac{d\overline{R}}{d\phi}=\langle f_3\rangle \overline{R}^3+\dots
\end{equation}
where 
$$
\langle f_3\rangle =\frac{1}{2\pi}\int_0^{2\pi}\frac{1}{d_0}
\left(-{\frac {c_{2}\,d_{1}}{d_{0}}} \right) d\phi~~.
$$
After some algebra we find
\begin{equation}\label{eq:f3}
\langle f_3\rangle=\,{\frac { \left( A_{110}+2\,B_{020} \right) A_{020}+A_{200}(A_{110}-2B_{200})-B_{110}\, \left( B_{020}+B_{200} \right) }{{8A
}^{2}}}
\end{equation}
Moreover, for the average value of the angle $\phi$ we get
\begin{equation}\label{eq:dphidtauave}
\frac{d\overline{\phi}}{d\tau}\simeq A\Rightarrow\overline{\phi}
=\phi_0+A\tau
\end{equation}
Thus we can write:
\begin{equation}\label{eq:drdtauave}
\frac{d\overline{R}}{d\tau}=\frac{d\overline{R}}{d\overline{\phi}}
\frac{d\overline{\phi}}{d\tau}=\langle f_3\rangle \overline{R}^3A
\end{equation}
which, upon integration, gives the average distance as a function 
of time:
\begin{equation}\label{eq:spiral}
\overline{R}(\tau)=
\frac{R_0}{\sqrt{1-2R_0^2\langle f_3\rangle A \tau}}
=\frac{R_0}{\sqrt{1-2R_0^2\langle f_3\rangle (\overline{\phi}
-\phi_0)}}~~.
\end{equation}
Equation (\ref{eq:spiral}) allows to see that the projection of the 
orbit on the F-plane ${\cal F}_0$ has the form of a spiral around 
the nodal point. In the forward sense of time, the spiral is inscribed 
counterclockwise (clockwise) if $A>0$ ($A<0$). Likewise, the nodal point 
acts as attractor, if $\langle f_3\rangle A<0$, or repellor, if 
$\langle f_3\rangle A<0$. Similarly to the 2-d case (see 
\cite{efthymiopoulos2007nodal}, \cite{Efth2009}), {\it Hopf 
bifurcations} of limit cycles can take place at critical instances 
(of parameterized time $\tau=\tau_c$, or $t=t_c$ in the original time) when 
$\langle f_3\rangle$ becomes equal to zero. Such limit cycles are 
generated at the nodal point, and they grow in radius away from, 
and surrounding, the nodal point, for times close to $t_c$.

We note,
that in  special models such as the one of \cite{wisniacki2005motion, borondo2009dynamical},
the Bohmian flow takes a conservative form which implies that the node
acts as a center and $\langle f_3\rangle$ is always equal to zero.
In this case, as established in  \cite{wisniacki2005motion, borondo2009dynamical},
one finds chaos close to the nodes due to the existence of a homoclinic tangle. An interesting question regards the connection of homoclinic theory with the mechanism of chaos discussed in \cite{Efth2009} and in the present paper, which is based on the scattering of Bohmian trajectories close to the nodal points (see section III below).

\begin{figure}[ht]
\includegraphics[scale=0.65]{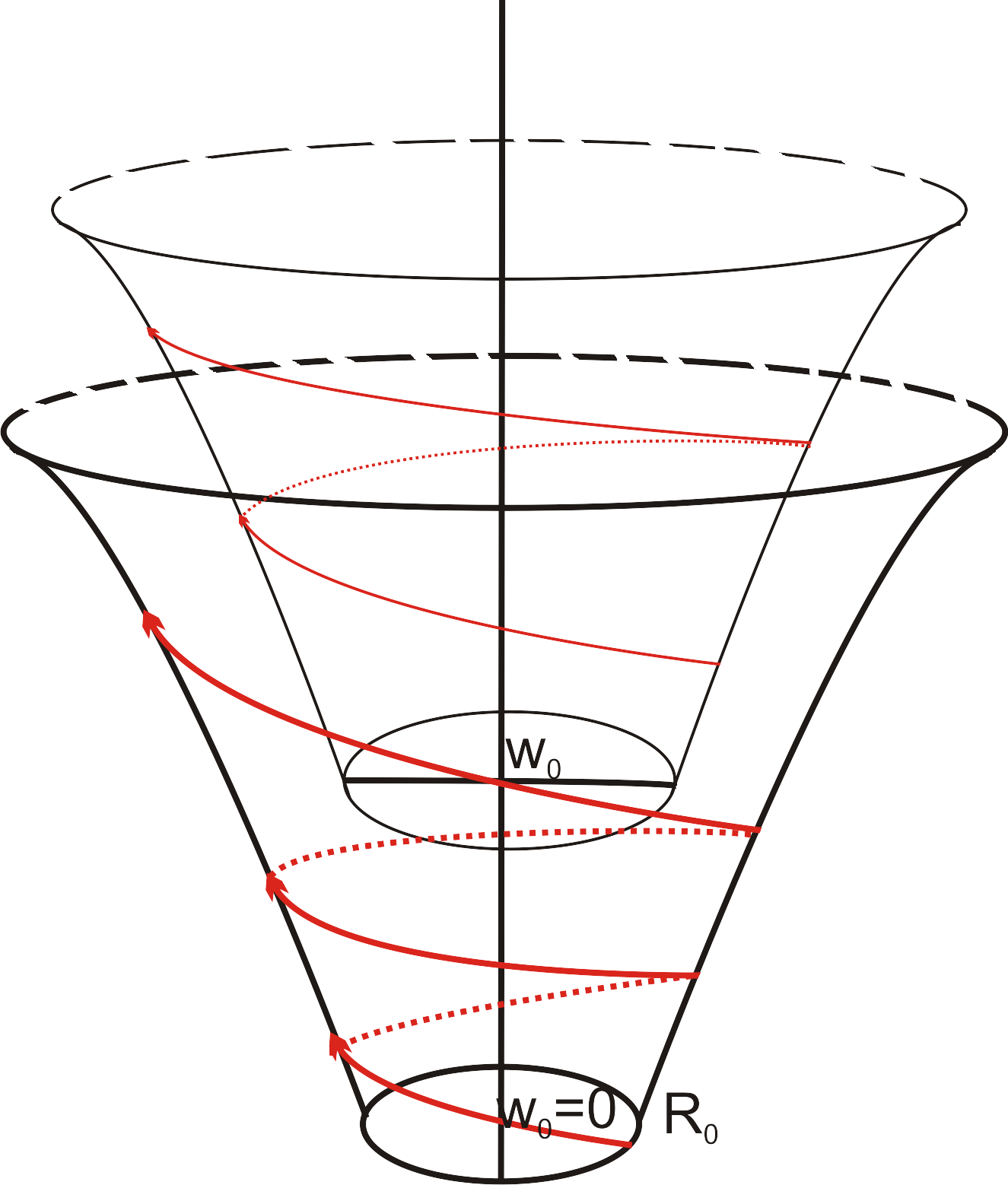}
\caption{Schematic representation of the trajectories in the neighborhood of 
one nodal point along a nodal line. The trajectories are helical and they lie 
in a foliation of conoidal surfaces of revolution, generated by the envelope
curves given by Eq.(\ref{eq:wlogr}) and parameterized by $(R_0, w_0)$.}
\label{fig:helix}
\end{figure}

The evolution in the third dimension, normal to the F-plane 
${\cal F}_0$ can now be approximated by substituting the `averaged' 
solutions $\overline{R}(\tau)$, $\overline{\phi}(\tau)$ of 
Eqs.(\ref{eq:spiral}) and (\ref{eq:dphidtauave}) into 
Eq.(\ref{eq:cylw}). Then
\begin{eqnarray}\label{eq:wtau}
w(\tau)=w_0 &+&
\exp\left(\int_0^{\tau}d\tau'k_2
\left(\overline{\phi}(\tau')\right)\overline{R}(\tau')\right)\\
&\times&\int_0^\tau d\tau'
\Bigg[e_2\left(\overline{\phi}(\tau')\right)\overline{R}^2(\tau')
\left(\exp\left(-\int_0^{\tau'}d\tau''k_2
\left(\overline{\phi}(\tau'')\right)
\overline{R}(\tau'')\right)\right)\Bigg]~~.\nonumber
\end{eqnarray}
The full 3-d geometric shape of the trajectories is helical, 
as one can see from the averaged equation
$$
\frac{d\overline{w}}{d\phi} = \frac{1}{ 2\pi}\int_0^{2\pi}
\left(\frac{e_2(\phi)R^2+k_2(\phi)R w+\ldots}{
d_0+d_1(\phi)R+h_1w+\dots}\right)d\phi~~.
$$
To the lowest approximation we find
$$
\frac{d\overline{w}}{d\overline{R}}=
\frac{d\overline{w}}{d\overline{\phi}}
\frac{d\overline{\phi}}{d\overline{R}}\approx 
\frac{\langle e_2\rangle}{ 
\langle f_3\rangle A \overline{R}}
$$
with solution
\begin{equation}\label{eq:wlogr}
\overline{w}(\overline{R})\approx w_0+ \frac{\langle e_2\rangle}{ 
\langle f_3\rangle A}\ln(\overline{R}/R_0)~~.
\end{equation}
Thus, a trajectory started at $(R_0,\phi_0,w_0)$ inscribes a 
helix, keeping a spiral projection in the F-plane ${\cal F}_0$ 
given by Eq.(\ref{eq:spiral}), while unfolding in the direction 
normal to ${\cal F}_0$ according to the logarithmic envelope 
$\overline{w}(\overline{R})$, given by Eq.(\ref{eq:wlogr}) 
(see Fig.\ref{fig:helix}).

 Note that the two equations taken 
together, (\ref{eq:spiral}) and (\ref{eq:wlogr}), define a 
foliation of cylindrical surfaces of revolution in which lie 
all helical trajectories parameterized by the different initial 
conditions $(R_0,\phi_0,w_0)$. 

\subsection{X-point}
\label{subsection:xpoint}
An `X-point' is defined as a critical point of the local quantum flow 
around a given nodal point, where the velocity of the trajectory in the 
inertial frame is equal to that of the nodal point. We find such points 
numerically using standard root-finding techniques. On the other hand, 
the existence, uniqueness (i.e. that only one X-point can be associated 
with the flow around each individual nodal point) and hyperbolic character (one real 
positive and one real negative eigenvalue of the linearized flow around 
the X-point) can be established in a similar way as in the 2-D case via 
the following approximations. We first require that the nodal 
point moves sufficiently fast so that the following inequality 
holds:
\begin{equation}\label{eq:vfast}
\max\{|V_u|,|V_v|\}>>
\frac{\max\{|a_{ijk}|,|b_{ijk}|,i+j+k=2\}}
{\min\{|a_{ijk}|,|b_{ijk}|,i+j=1, k=0\}}.
\end{equation}
The condition (\ref{eq:vfast}) implies that the theory does not apply
 if the nodal point is stationary or moves slowly. As explained in detail
 below, this means that the  scattering of the Bohmian trajectories,
 which leads to chaos, is a generic consequence of the fast motion of the
 nodal points. Thus, while counter examples can be found (see \cite{cesa2016chaotic}), the
 generic mechanism leading to chaos is connected with moving nodes in the
 3-d configuration space. Eqs.(\ref{eq:bohmexp}) yield
\begin{equation}\label{eq:dsquad}
\frac{du}{d\tau}= -Av - G_2V_u + F_{u,2},~~~
\frac{dv}{d\tau}= Au - G_2V_v + F_{v,2},~~~
\frac{dw}{d\tau}= F_{w,2}
\end{equation}
where $G_2=({a_{100}u+a_{010}v})^{2}+(b_{100}u+b_{010}v)^2$, while in view 
of the condition (\ref{eq:vfast}),
the terms $F_{u,2}$, $F_{v,2}, F_{w,2}$ can be disregarded as they
 are smaller in size than the maximum size 
of the terms $G_2V_u$, $G_2 V_v$. For a second stationary point of the 
flow, denoted $(u_X,v_X,w_X)$, the r.h.s. of Eqs.(\ref{eq:dsquad}) must 
become equal to zero. Then, taking into account Eq.(\ref{eq:dsquad}), 
the following relation holds:
\begin{equation}\label{eq:sx}
s_X\equiv\frac{v_X}{u_X}\simeq -\frac{V_u}{V_v}.
\end{equation}
The approximation (\ref{eq:sx}) can be used to obtain a sequence 
of approximants $(u_X^{(r)},v_X^{(r)},w_X^{(r)})$ of the exact 
stationary point as follows: replacing first Eq.(\ref{eq:sx}) into 
the last of Eqs.(\ref{eq:bohmexp}) with $\frac{dw}{dt}=0$ 
yields a first approximant:
$$
w_X^{(1)} = C_X u_X^{(1)}
$$
with 
$$
C_X=\left(-\frac{C_{200}+C_{020}s_X^2+C_{110}s_X}
{C_{101}+s_XC_{011}}\right)~~.
$$
Substitution to the first and second of Eqs.(\ref{eq:bohmexp}) 
yields a {\it unique} non-zero first approximant for $u_X$ and 
$v_X$:
\begin{equation}
\nonumber u_{x}^{(1)}=-\frac {A}{B_x},\quad v_{x}^{(1)}=\frac {A}{A_x}
\end{equation}
with
$$
A_X=A_{020}+\frac{1}{s_X}\left(A_{110}+A_{011}C_X\right)
+\frac{1}{s_X^2}\left(A_{200}+A_{002}C_X^2\right)
$$
$$
B_X=B_{020}s_X^2+B_{110} s_X+B_{200}
+B_{002}C_X^2 + B_{101}C_X~~.
$$
Starting with the approximant $(u_X^{(1)},v_X^{(1)},w_X^{(1)})$ 
we can finally refine the solution for the stationary point, e.g.  via 
successive Newton iterates. The key remark is that, due to (\ref{eq:vfast}),
one finds that $R_X=\sqrt{u_X^2+v_X^2}={\cal O}(V^{-1})$, where 
$V=\sqrt{V_u^2+V_v^2}$. Thus, the X-point remains in the neigbourhood of 
the nodal point only when the nodal point moves fast.

The hyperbolic character of the stationary `X-point' $(u_X,v_X,w_X)$ 
can be established as follows:
The equations of the flow (\ref{eq:bohmexp}) come from the general
expression:
\begin{equation}
\Big(\frac{du}{d\tau},\frac{dv}{d\tau},\frac{dw}{d\tau}\Big)^T
=\rho\nabla S,\label{genflow}
\end{equation}
where $\rho\equiv G=\Psi\Psi^*, S=\arctan[\Im(\Psi)/\Re(\Psi)]$.
Setting $u=u_X+\xi, v=v_X+\eta, w=w_x+\zeta$ and linearizing 
Eqs.~(\ref{genflow}) we obtain:
\begin{equation}\label{eq:varx}
\Big(\frac{d\xi}{d\tau},\frac{d\eta}{d\tau},\frac{d\zeta}{d\tau}\Big)^T=M\cdot (\xi, \eta, \zeta)^T,
\end{equation}
where $M$ is the $3\times 3$ variational matrix with elements
given by:
$$
M_{ij}=\left(\rho\frac{\partial^2 S}{\partial q_iq_j}\right)\Bigg|_
{q_1=u_X,q_2=v_X,q_3=w_X},~~~i,j=1,2,3
$$
 and 
$q_1\equiv u$, $q_2\equiv v$, $q_3\equiv w$.
The matrix $M$ is 
symmetric, therefore all three eigenvalues $\lambda_i$, $i=1,2,3$, 
are real. Using now the expansion (\ref{eq:bohmexp}), we find
\begin{equation}\label{eq:varmat}
M=\frac{1}{G_X}
\left(\begin{tabular}{ccc}
$A v_X\frac{\partial G_X}{\partial u_X}+{\cal O}_3$ 
&$-AG_X +Av_X\frac{\partial G_X}{\partial v_X}+{\cal O}_3$ 
&$~~~~~~{\cal O}_3$\\
$AG_X-Au_X\frac{\partial G_X}{\partial u_X}+{\cal O}_3$
&$-Au_X\frac{\partial G_X}{\partial v_X}+{\cal O}_3$ 
&$~~~~~~{\cal O}_3$\\
${\cal O}_3$ 
&${\cal O}_3$ 
&$~~~~~~{\cal O}_3$
\end{tabular}\right)
\end{equation}
where $G_X = G_2(u_X,v_X)$, and ${\cal O}_3$ denotes terms of order 3 
in the variables $(u_X,v_X,w_X)$ (while only second order terms 
are explicitly written in Eq.(\ref{eq:varmat})). One can easily check
that $-AG_X+Av_x\frac{\partial G_X}{\partial v_X}=
AG_X-Au_X\frac{\partial G_X}{\partial u_X}$, i.e. the matrix in the 
r.h.s. of Eqs~(\ref{eq:varmat}) preserves the symmetry of the full 
matrix $M$ to the leading order. We then find: 
\begin{equation}\label{eq:eigx}
\lambda_1\lambda_2 = -A^2 + {\cal O}_{1},~~ \lambda_3={\cal O}_1~~.
\end{equation}
This implies that the linear stability character of the X-point 
is hyperbolic, containing at least one stable and one unstable 
eigendirections, while $|\lambda_3|<<\min\{|\lambda_1|,|\lambda_2|\}$, 
provided that the distance $d_X$ of the X-point from the nodal point 
is small. In fact, from the equations for the first approximant 
$(u_x^{(1)},v_x^{(1)},w_x^{(1)})$ we find that $d_X={\cal O}(V^{-1})$, 
where $V$ is the velocity of the nodal point $V=(V_u^2+V_v^2)^{1/2}$.

\subsection{X-line and construction of the 3-d quantum vortex}
\label{subsec:npxpc}
According to the analysis of the previous subsections, to every nodal 
point $\mathbf{r}_0$ along a nodal line we can  associate a single 
X-point with the properties mentioned in subsection \ref{subsection:xpoint}. 
Using the notation of subsection \ref{subsection:nodline}, the position of 
the nodal point $\mathbf{r}_0$ is parameterized in terms of the length 
parameter $s$, starting from an arbitrary `origin' O in the nodal line, 
through the continuous vector function 
$\mathbf{r}_0(s)\equiv(x_0(s),y_0(s),z_0(s))$. Let 
$(u_X(s),v_X(s),w_X(s))$ be the local co-ordinates of the X-point 
associated with the nodal point $\mathbf{r}_0(s)$. Assuming, 
now, the co-ordinate representation (\ref{eq:frenet}) is adopted, 
the curve defined by the parametric relations:
\begin{equation}\label{eq:xlinepar}
\mathbf{r}_X(s)=\left(
\begin{tabular}{c}
$x_X(s)$\\
$y_X(s)$\\
$z_X(s)$
\end{tabular}
\right)=
\left(
\begin{tabular}{c}
$x_0(s)$\\
$y_0(s)$\\
$z_0(s)$
\end{tabular}
\right)+
\left(
\begin{tabular}{ccc}
$n_x$ &$n_y$ &$n_z$ \\
$b_x$ &$b_y$ &$b_z$ \\
$t_x$ &$t_y$ &$t_z$
\end{tabular}
\right)^{-1}
\left(
\begin{tabular}{c}
$u_X(s)$\\
$v_X(s)$\\
$w_X(s)$
\end{tabular}
\right)
\end{equation}
where $(n_x,n_y,n_z)$, $(b_x,b_y,b_z)$, $(t_x,t_y,t_z)$ are the 
xyz-coordinates of the normal, bi-normal and tangent vectors $\mathbf{n}(s)$, 
$\mathbf{b}(s)$, $\mathbf{t}(s)$, is the `X-line' associated with the nodal 
line $\mathbf{r}_0(s)$. 

It is to be emphasized that, since the dynamical system 
(\ref{eq:bohm3b}) was defined around {\it a single} nodal point 
along the nodal line $\mathbf{r}_0(s)$ (i.e. for one value of 
the parameter $s$), the X-line $\mathbf{r}_X(s)$ defined as in 
Eq.(\ref{eq:xlinepar})  gives a {\it unique invariant hyperbolic point}  
corresponding to the fixed value of $s$. However, all other points of 
the X-line {\it do not} form an invariant set in this particular co-ordinate system
$(u, v, w)$. On the other hand, as shown in 
Appendix I, starting from the whole family of dynamical systems 
of the form (\ref{eq:bohm3b}), it is possible to construct a 
unique dynamical system in which the X-line becomes an exact, 
normally hyperbolic, invariant manifold of the flow. In this new system, 
each point along the X-line is itself invariant, thus the stable 
and unstable manifolds of the X-line are formed by the union of 
the stable and unstable manifolds of each of the X-points along 
the X-line. 

\begin{figure}
\includegraphics[scale=0.6]{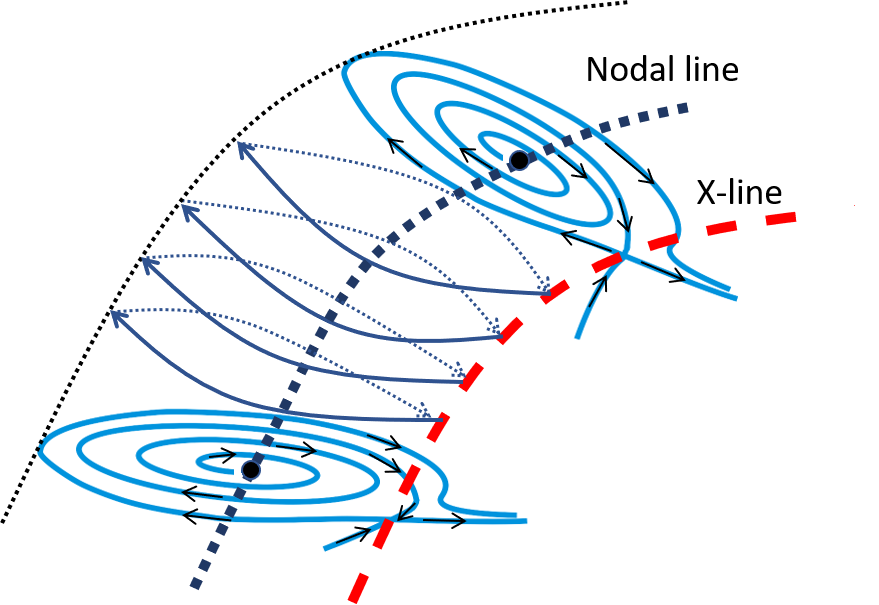}
\caption{Schematic illustration of a case in which one of the manifold branches
from  the X-line (foliation of all blue curves) terminates, after spiral-like 
revolution, at   the nodal line (square dotted line in the center of the foliation).}
\label{hopfI}
\end{figure}

\begin{figure}
\includegraphics[scale=0.6]{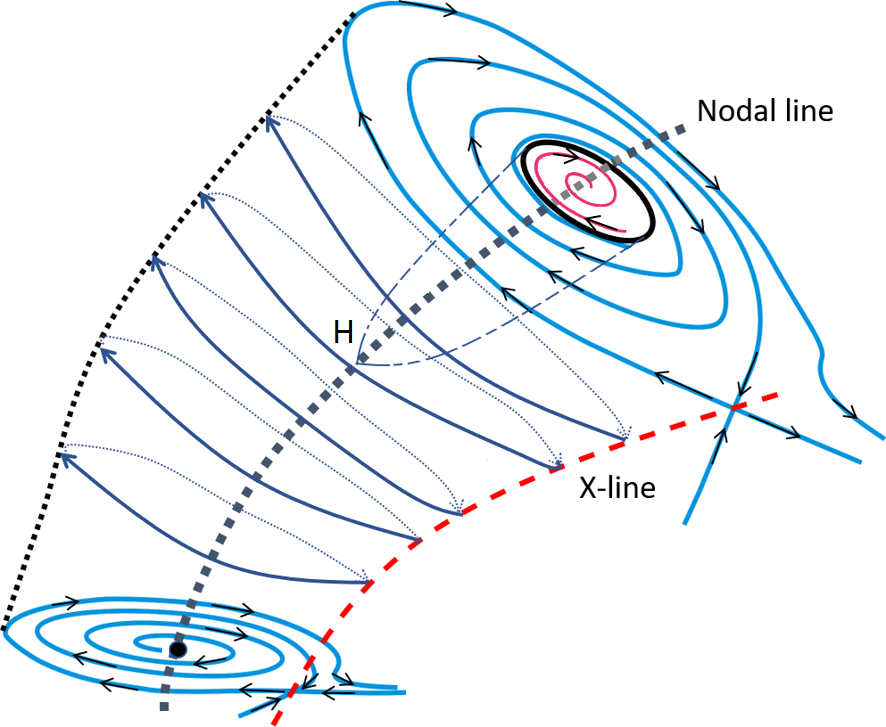}
\caption{Schematic illustration of  a Hopf bifurcation  taking place at
the nodal point $H$ of the nodal line (square dotted line in the center of the foliation). The bold closed curve in the upper section of the plot 
represents a limit cycle.  Trajectories starting both outside or inside the limit
cycle tend asymptotically to the latter in the forward or backward sense of time,
depending on whether the limit cycle is attractor or repellor.}\label{hopfII}
\end{figure}

Joining this analysis with the local analysis around nodal points, 
we arrive at the following picture, shown schematically in 
Figs.~\ref{hopfI} and \ref{hopfII}.
The X-line's stable and unstable manifolds (${\cal W}_X^S$ and 
${\cal W}_X^U$ respectively) are two-dimensional surfaces 
passing through the X-line. Depending on the values of the wavefunction coefficients 
(which change in time), one of the branches of either 
${\cal W}_X^S$ or ${\cal W}_X^U$ continues as a spirally-twisting 
cylindrical surface which terminates at a second invariant set of 
the flow. In the simplest case, shown in Fig.~\ref{hopfI}, the manifold 
terminates in the nodal line. However, as noted in subsection 
\ref{subsec:vortices}, Hopf bifurcations of limit cycles take 
place at every change of sign of the parameter $\langle f_3\rangle$ in 
Eq.(\ref{eq:drdphif3}). The value of $\langle f_3\rangle$ depends on the nodal 
point $\mathbf{r}_0$ around which the wavefunction is expanded. 
Thus, Hopf bifurcations are possible to occur in time, for a 
specific moving nodal point, or in space, i.e., as we consider 
different nodal points in the nodal line at a fixed time $t$. 
Fig.~\ref{hopfII} shows schematically the form of the cylindrical structure 
of NPXPCs in this latter case: the manifolds of some X-points are 
spirals terminating at a corresponding nodal point, while, for 
other X-points, the manifolds form spirals terminating at a 
limit cycle, while the spiral flow has opposite pitch angle 
inside each limit cycle.

Finally, the three remaining branches of the invariant manifolds 
of the X-line, which do not evolve as spirals, necessarily follow 
the nearby flow. One branch surrounds the whole cylindrical 
structure of NPXPCs, and then goes to infinity, while the 
remaining two branches extend from the start to infinity. 

Precise  computations 
of such complexes are hardly tractable, since explicit formulas 
for the associated dynamical system, in which ${\cal W}_X^S$ and 
${\cal W}_X^U$ are to be constructed (see Appendix I), can only 
be given in approximative form. In practice, however, we can 
still use the approximation of a point-by-point approximation 
of these manifolds via the use of Eqs.(\ref{eq:bohm3b}) computed 
at different nodal points. Examples of this form are given in 
the next section. 

Finally, we stress again that all critical sets and asymptotic 
manifolds defined in this and in previous subsections are 
invariants of the quantum flow only under the `frozen' approach 
to the equations of motion (Eqs.\ref{eq:bohm3aut}), i.e., only 
under the assumption that an adiabatic approximation holds, 
and that all Bohmian velocities within the NPXPCs are larger 
than the velocity $V$ of the nodal point itself. Returning to 
the equations in the original time $t$ without any approximation (Eqs.~\ref{eq:bohm3}), 
at points of distance $d$ from a nodal point the velocities 
are of order ${\cal O}(d^{-1})$. One thus gets that the adiabatic 
approximation holds in regions of size $d<V^{-1}$. Since the 
X-point is at a distance $d_X\sim V^{-1}$, the adiabatic 
approximation holds for $d<d_X$. Hence, the outer limit of the 
cylindrical structure of NPXPCs marks at the same time the 
domain of validity of the adiabatic approximation. On the other 
hand, since the X-line is located at the border of this domain, 
the identification of the X-line as the main mechanism of 
hyperbolic scattering of the trajectories necessitates numerical 
validation. Such validation is provided by specific 
numerical examples in the section to follow.  

\section{Numerical application}
\label{sec:num}

In this section we make a numerical application of the theory in the 
case of the wave function of 3-d harmonic oscillator:
\begin{align}\label{form}
\Psi(\vec{x},t)=&a\Psi_{p_1,p_2,p_3}(\vec{x},t)
+b\Psi_{r_1,r_2,r_3}(\vec{x},t)
+c\Psi_{s_1,s_2,s_3}(\vec{x},t),
\end{align}
where $
\Psi_{n_1,n_2,n_3}(\vec{x},t)
=\Psi_{n_1,n_2,n_3}(\vec{x})e^{-iE_it/\hbar}$
and $\Psi_{n_1,n_2,n_3}(\vec{x})$ are eigenstates of the 3-d 
harmonic oscillator of the form
\begin{align}\label{eigenstate}
\Psi_{n_1,n_2,n_3}(\vec{x})=
\prod_{k=1}^3\frac{\Big(\frac{m_k\omega_k}{\hbar\pi}\Big)^{\frac{1}{4}}
\exp\Big(\frac{-m_k\omega_kx_k^2}{2\hbar}\Big)}{\sqrt{2^{n_k}n_k!}}
H_{n_k}\Big(\sqrt{\frac{m_k\omega_k}\hbar}x_k\Big),
\end{align}
$n_1,n_2,n_3$ stand for their quantum numbers, $\omega_1,\omega_2,\omega_3$ 
for their frequencies and $E_1,E_2,E_3$ for their energies. Hereafter we 
set $m_i=\hbar=1$ and write $x_1,x_2,x_3$ as $x,y,z$. For a given triplet 
of quantum numbers $(n_1,n_2,n_3)$ we have 
$E=\sum_{i=1}^3(n_i+\frac{1}{2})\omega_i$. 

\begin{figure}
\includegraphics[scale=0.4]{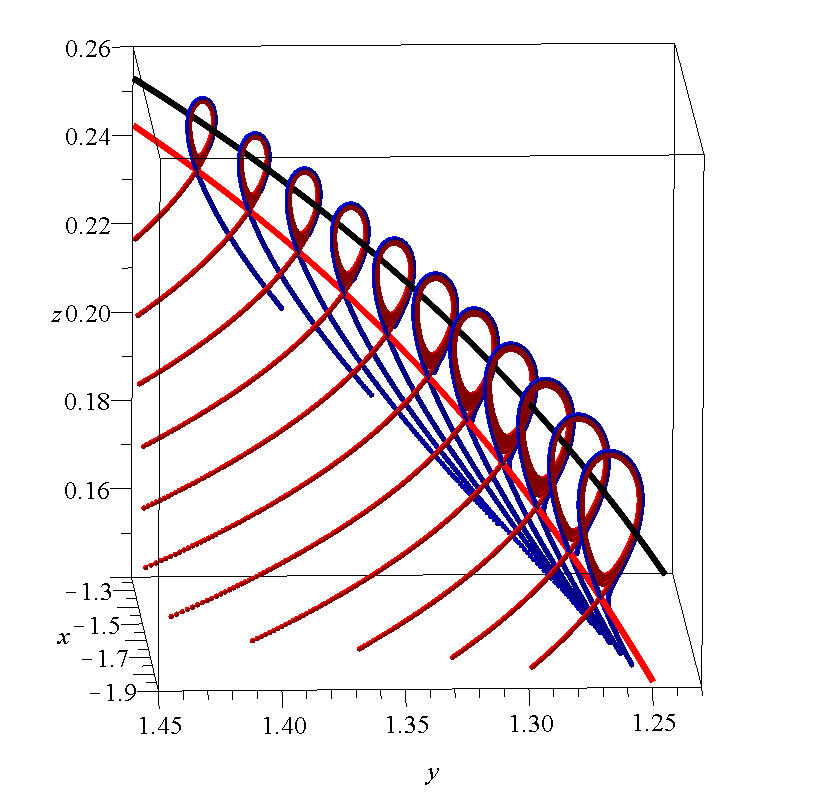}
\caption{Full numerically computed (non-schematic) example of
the local quantum flow around a segment of a nodal line, leading
to the formation of a 3-d structure of NPXPCs, in the case
of the 3-d harmonic oscillator system described by Eq.~(\ref{sistima}) at the time $t=4$. The black curve represents the nodal line. 
The red curve is the X-line, which connects the X-points of all 
complexes. The four branches of the stable (blue) and
unstable (red) manifolds of the X-line are also plotted.
One branch of the unstable manifold forms spirals surrounding and tending asymptotically to the nodal line. }\label{full}
\end{figure}

\begin{figure}[ht]
\includegraphics[scale=0.35]{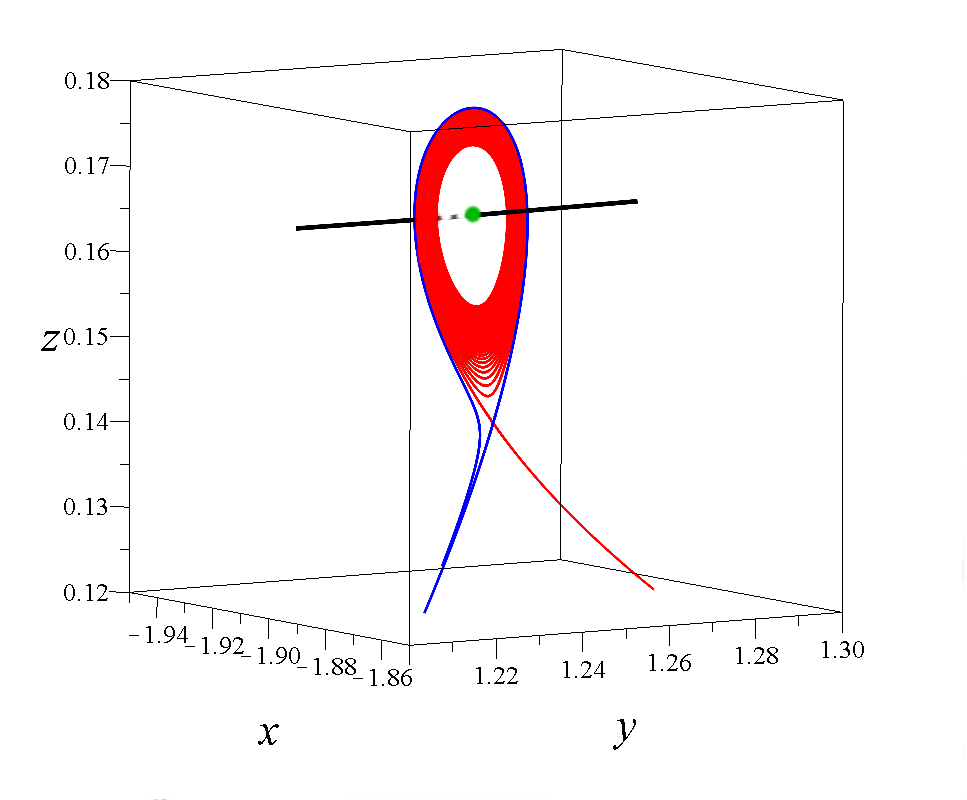}
\caption{Magnification of a single nodal point-X-point complex of the 3-d nodal 
point-X-point structure shown in Fig.~\ref{full}.}\label{single}
\end{figure}

Chaotic trajectories scattered by the X-lines formed near moving nodal lines are 
typically found in systems of the form (\ref{form}) (see, for example \cite{Tzemos2016, contopoulos2017partial}). As an application of the theory of section \ref{sec:genthe}
we now focus on the case  $a=b=c=1/\sqrt{3}$, 
$p_1=p_2=p_3=0$, $r_1=r_3=1, r_2=0$, $s_1=0, s_2=1, s_3=2$, namely on the 
wavefunction
\begin{align}
\Psi(\vec{x},t)=\frac{1}{\sqrt{3}}\Big(\Psi_{0,0,0}(\vec{x},t)
+\Psi_{1,0,1}(\vec{x},t)+\Psi_{0,1,2}(\vec{x},t)\Big).\label{sistima}
\end{align}
Figure (\ref{full}) shows a real (non-schematic) computation of the 3-d cylindrical
structure of NPXPCs in the system at the time $t=4$. 
We observe the multiple nodal point-X-point complexes along the nodal 
line (black curve) and the red X-line which is the connection of 
X-points. Figure (\ref{single})  shows a detail of Fig~(\ref{full}), i.e. 
 a single nodal point X-point 
complex clearly depicting  the invariant manifolds  emanating from the 
X-point, and in particular the spiral motion around the nodal point. 

The chaotic scattering effects in the neighborhood of the 3-d cylindrical
structure of NPXPCs can be unravelled numerically as follows: 
in order to quantify chaos, we use as an indicator 
the local Lyapunov characteristic number, or `stretching number' \cite{voglis1994invariant, Contopoulos200210}. If $\xi_k$ is the length of 
the deviation vector between two nearby trajectories at the time 
$t = \kappa t_0,\, \kappa = 1, 2,\dots$,
the  stretching number is defined as
\begin{equation}
\alpha_\kappa=\ln\Big(\frac{\xi_{\kappa+1}}{\xi_\kappa}\Big)\label{str}
\end{equation}
and the ``finite time Lyapunov characteristic number" is given by the equation:
\begin{equation}
\chi=\frac{1}{\kappa t_0}\sum_{i=1}^\kappa\ln\alpha_i\label{flcn}
\end{equation}
The LCN is the limit of $\chi$ when $\kappa\to\infty$.

\begin{figure}[ht]
\includegraphics[scale=0.7]{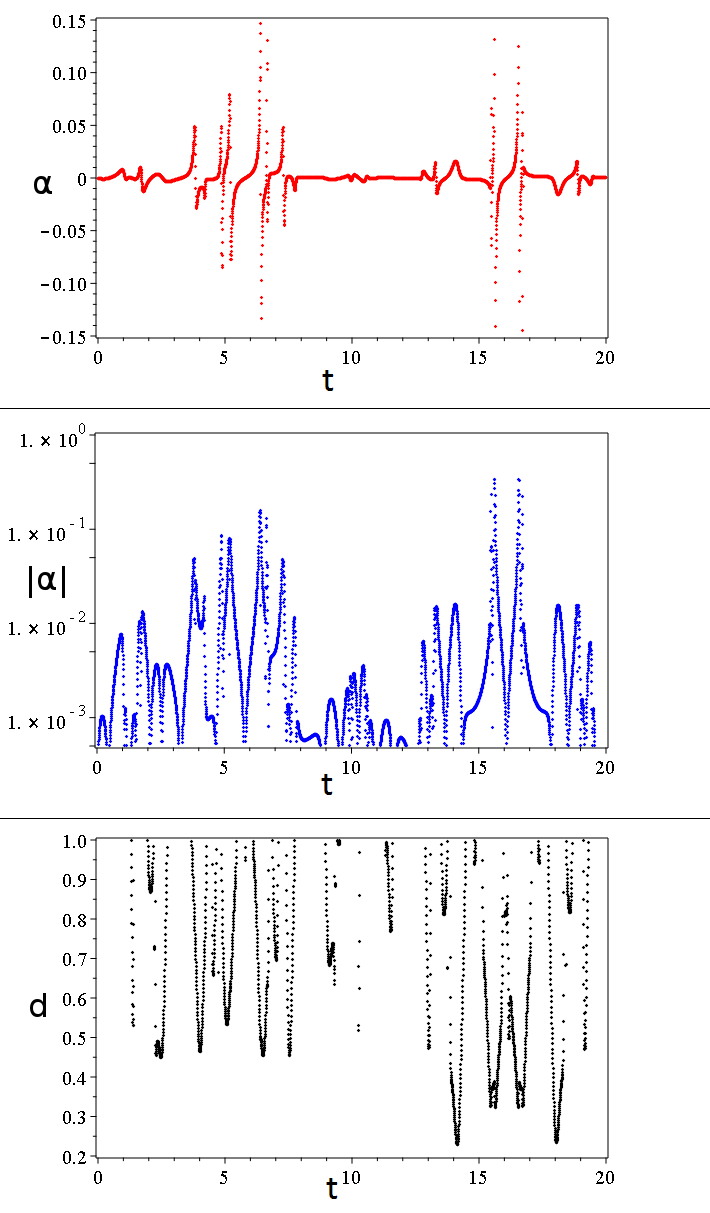}
\caption{Multiple scattering events for a Bohmian trajectory. The time evolution of the `stretching number' $\alpha$ (upper panel), absolute value $|\alpha|$ in logarithmic scale (middle panel), and minimum distance from the X-line (lower panel) are shown in the time interval $0\leq t\leq 20$. The trajectory is computed in the model (\ref{form}), with wavefunction (\ref{sistima}), and initial conditions $x(0)=-0.7, y(0)=-1.1, z(0)=1.3$. }\label{pliris}
\end{figure}

Figure \ref{pliris} shows a  numerical investigation of the chaotic scattering effects
close to the nodal line based on the use of Eqs.(\ref{str}) and (\ref{flcn}). The example shows 
 a Bohmian trajectory in the system (\ref{sistima}) for $t\in[0,20]$, for which we compute the time evolution of the stretching number $\alpha$. The upper
(red) curve shows shows $\alpha$ as a function of time, while the middle (blue)
curve shows the absolute value $|\alpha|$ in a logarithmic scale, allowing to show more clearly weak
scattering events. Finally the lower (black) 
curve shows the minimum distance between the trajectory and the X-line at every time $t\in[0,20]$. This is done by 
projecting the trajectory to the closest nodal point of the nodal line 
corresponding to a given time $t$, and then calculating the X-point of the 
corresponding F-plane.
This is an approximative method that simplifies the calculations, while still producing the key result with a satisfactory accuracy. Whenever a Bohmian
trajectory comes close to one X-point of a 3-d structure of NPXPCs it gets scattered
by it. At the same time the local Lyapunov exponent undergoes a jump and chaos is produced.
The main effect, which depicts the cumulative mechanism of generation of chaos for 
the 3-d trajectory, is unraveled by comparing the times when jumps in 
the values of the stretching number appear with the times when the 
trajectory has its closest approaches to the X-line. These times 
appear to always coincide, while the largest jumps take place at distances 
$d<<1$. In fact, a careful inspection of the profile of the time evolution 
of the stretching number at every jump reveals profiles analogous in 
shape to those characterized as `type I' or `type II' in chaotic trajectories 
in the 2-d case (see \cite{Efth2009}). Also, no jumps are observed when the trajectories 
are at distances $d>1$ from the X-line. This phenomenon is confirmed by 
repeating the computation of the local Lyapunov characteristic numbers in 
various trajectories in the same system, or in systems with different 
wavefunctions.


\section{Conclusions}
\label{sec:concl}
In this paper we provide a general theory allowing to interpret the 
appearance of complex, or chaotic, behavior for quantum (Bohmian) trajectories 
tracing the quantum currents in 3-dimensional quantum systems. This theory 
extends results found in \cite{Efth2009} from the 2-d to the 3-d case, and it is 
based on general formulas which allow to characterize the form of the 
quantum flow in the vicinity of 3-d quantum vortices. In particular:

\begin{enumerate}
\item{Starting from the analysis of \cite{falsaperla2003motion},
 where it is shown that in the 
close neighborhood of a nodal line the quantum flow becomes stratified 
in planes orthogonal to the nodal line (here called the `F-planes'), 
we provide an analysis of the nonlinear terms in the equations of motion 
of trajectories tracing the quantum currents, as viewed in a frame of 
reference locally co-moving with every node along the nodal line. 
This analysis yields formulas derived by a generic second-order expansion 
of the wavefunction around nodal points, thus it applies to arbitrary 
3-d wavefunctions $\Psi$.}
\item{As a consequence of symmetries identified in the transformed second
order equations (which reflect the positiveness and preservation of  
quantum probability), we show that in the neighborhood of a time-evolving 
nodal line the quantum flow is spiral-helical, i.e., it describes 
spirals in the F-planes while drifting in a direction parallel to the 
nodal line. Approximative formulas are given which describe both the 
spiral motion and the drift. In particular, we define a crucial coefficient ($\langle f_3\rangle$, see Eq.(\ref{eq:f3})), whose 
zeros describe Hopf bifurcations taking place both in time (the nodal line deforms as a whole), or in space (i.e. along a nodal 
line at a fixed time $t$). Such bifurcations give rise to limit cylinders, which generalize 
the phenomenon of the bifurcation of limit cycles observed in \cite{efthymiopoulos2007nodal}.}
\item{Further away from a nodal line, we establish the existence of 
a normally hyperbolic one-dimensional curve called the `X-line', whose three branches of 
manifolds extend to infinity, while one branch approaches asymptotically the 
nodal line by  spirally revolving around it. The overall 
geometry of the quantum flow formed by the nodal line, the X-line 
and the latter's asymptotic manifolds defines a `3-d structure 
of nodal point - X-point complexes'. We provide both schematic 
and real (numerical) examples of computation of such 3-d structures, 
and we examine their precise form depending on whether or not Hopf 
bifurcations take place along the nodal line. Finally, we argue that  motion of the nodal line is a necessary condition for the X-line 
to be formed, as well as for the latter's hyperbolic character.}
\item{We finally study the emergence of chaos in examples of 
trajectories computed fully numerically, i.e. without any analytical approximation 
of the equations of motion. We establish that the sensitivity to the 
initial conditions, quantified by the accumulation of positive values 
of the local Lyapunov Characteristic Numbers, is strictly correlated with 
close encounters of the trajectories with the X-line formed around 
every moving nodal line. The emergence of chaos 
through multiple scattering events of the trajectories
 with the X-lines is a generically 
observed behavior, in the sense that it appears for different initial conditions 
of the  trajectories in the same quantum system,
but also in systems with different (arbitrary) choice of 3-d 
wavefunction.}
\end{enumerate}

\begin{acknowledgments}
This research was supported by the Research Committee of the Academy of Athens.
\end{acknowledgments}
\clearpage
\section{Appendix 1}

Consider a segment ${\cal S}$ of a nodal line $\mathbf{r}_0(s)$ and a 
domain around ${\cal S}$ defined by: 
\begin{equation}\label{eq:domainv}
{\cal U}=\cup_{\mbox{all $s: \mathbf{r}_0(s)\in {\cal S}$}}{\cal V}_{\mathbf{r}_0(s),\epsilon(s)} 
\end{equation}
where ${\cal V}_{\mathbf{r}_0(s),\epsilon(s)}$ denotes the domain 
${\cal V}_\epsilon$ defined in subsection \ref{subsec:vortices} for 
the nodal point corresponding to a specific value of $s$. Assume that the 
radii of curvature drawn from every point of ${\cal S}$ do not intersect 
within ${\cal U}$. A curvilinear transformation $(x,y,z)\rightarrow(U,V,S)$ 
can be defined within ${\cal U}$ as follows: For every point $P\equiv(x,y,z)
\in{\cal U}$: if $P\in{\cal S}$, set $S=s$, $U=V=0$, where $s$ is the 
parameter value corresponding to P. If, now, $P\notin{\cal S}$, draw 
the normal from $P$ to ${\cal S}$, which intersects ${\cal S}$ at a 
certain point with parameter value $s$. Then, set $S=s$. Compute the 
normal and bi-normal vectors $\mathbf{n}(s)$, $\mathbf{b}(s)$, and 
set $U = xn_x+yn_y+zn_z$, $V=xb_x+yb_y+zb_z$. 

We denote the above transformation as:
$$
U=f_U(x,y,z),~~V=f_V(x,y,z),~~S=f_S(x,y,z)~~.
$$
Due to our adopted assumptions and domains, the functions $f_U,f_V,f_S$ 
are smooth and invertible. The inverse functions are denoted by 
$x=f_x(U,V,S)$, $y=f_y(U,V,S)$, $z=f_z(U,V,S)$. 

Take the complete Bohmian equations in the {\it inertial} frame 
(Eqs.(\ref{eq:bohm})), and use the chain rule on the functions $f_U, 
f_V,f_S$ and their inverses $f_x,f_y,f_z$. This allows to find the 
the form that the equations (\ref{eq:bohm}) take in the new variables, 
resulting in the system:
\begin{equation}\label{eq:equvsiner}
\frac{dU}{dt}=F^{(inertial)}_U(U,V,S),~~ 
\frac{dV}{dt}=F^{(inertial)}_V(U,V,S),~~ 
\frac{dS}{dt}=F^{(inertial)}_S(U,V,S) 
\end{equation}
where the functions $F^{(inertial)}_U,F^{(inertial)}_V,F^{(inertial)}_S$ 
are computed by applying the chain rule to both sides of Eq.(\ref{eq:bohm}). 

Consider, now, a certain nodal point $\mathbf{r}_0(s)$ and compute its 
associated X-point $\mathbf{r}_x(s)$ (Eq.(\ref{eq:xlinepar})). This is 
a new point with co-ordinates $x_X,y_X,z_X$. Use the transformation 
of the previous paragraph, and compute $(U_X,V_X,S_X)$. For every $s$ 
we have a unique X-point, thus a unique value $S_X$. Thus, we can define 
a function $s=\sigma(S_X)$, which we assume  to be smooth. Finally, 
the nodal point $\mathbf{r}_0(s)$ has a velocity $\mathbf{V}(s)$ defined 
by Eq.(\ref{eq:nodalvel}). 

The following proposition is an immediate consequence of the above 
definitions: the dynamical system defined by
\begin{eqnarray}\label{eq:fufvfs}
\frac{dU}{dt}&=&F^{(inertial)}_U(U,V,S)_{t=t_0}
-\mathbf{V}(\sigma(S))\cdot\mathbf{n}(\sigma(S))\nonumber\\
\frac{dV}{dt}&=&F^{(inertial)}_V(U,V,S)_{t=t_0} 
-\mathbf{V}(\sigma(S))\cdot\mathbf{b}(\sigma(S))\\
\frac{dS}{dt}&=&F^{(inertial)}_S(U,V,S)_{t=t_0} \nonumber
\end{eqnarray}
possesses an exactly invariant X-line, which coincides with the one given by 
Eq.(\ref{eq:xlinepar}). 

The quantities appearing in the r.h.s. of Eqs.(\ref{eq:fufvfs}) depend 
only on $U,V,S$, and they constitute the functions $F_U,F_V,F_V$ of 
Eq.(\ref{eq:eqmouvs}).  In fact, by the above construction we can see that 
the Jacobian matrix of the system (\ref{eq:fufvfs}) is symmetric at any point along the X-line, thus its eigenvalues are real. 
Since every point on the X-line is invariant, one eigenvalue of the 
Jacobian matrix is necessarily equal to zero, and the corresponding 
eigenvector is tangent to the X-line. On the other hand, taking any 
arbitrary point on the nodal line as the origin O, the 
transformation $(x,y,z)\rightarrow(U,V,S)$ differs from the one 
considered in subsection (\ref{subsec:vortices}), namely, 
$(x,y,z)\rightarrow(u,v,w)$, only by terms nonlinear in the quantities 
$U,V,S$. Thus, Eq.(\ref{eq:eigx}) still holds in the variables 
$(U,V,S)$. This implies that every point on the  X-line is 
hyperbolic, hence, the X-line is a normally hyperbolic one-dimensional 
invariant manifold.  

\bibliography{nbib}

\begin{thebibliography}{50}%
\makeatletter
\providecommand \@ifxundefined [1]{%
 \@ifx{#1\undefined}
}%
\providecommand \@ifnum [1]{%
 \ifnum #1\expandafter \@firstoftwo
 \else \expandafter \@secondoftwo
 \fi
}%
\providecommand \@ifx [1]{%
 \ifx #1\expandafter \@firstoftwo
 \else \expandafter \@secondoftwo
 \fi
}%
\providecommand \natexlab [1]{#1}%
\providecommand \enquote  [1]{``#1''}%
\providecommand \bibnamefont  [1]{#1}%
\providecommand \bibfnamefont [1]{#1}%
\providecommand \citenamefont [1]{#1}%
\providecommand \href@noop [0]{\@secondoftwo}%
\providecommand \href [0]{\begingroup \@sanitize@url \@href}%
\providecommand \@href[1]{\@@startlink{#1}\@@href}%
\providecommand \@@href[1]{\endgroup#1\@@endlink}%
\providecommand \@sanitize@url [0]{\catcode `\\12\catcode `\$12\catcode
  `\&12\catcode `\#12\catcode `\^12\catcode `\_12\catcode `\%12\relax}%
\providecommand \@@startlink[1]{}%
\providecommand \@@endlink[0]{}%
\providecommand \url  [0]{\begingroup\@sanitize@url \@url }%
\providecommand \@url [1]{\endgroup\@href {#1}{\urlprefix }}%
\providecommand \urlprefix  [0]{URL }%
\providecommand \Eprint [0]{\href }%
\providecommand \doibase [0]{http://dx.doi.org/}%
\providecommand \selectlanguage [0]{\@gobble}%
\providecommand \bibinfo  [0]{\@secondoftwo}%
\providecommand \bibfield  [0]{\@secondoftwo}%
\providecommand \translation [1]{[#1]}%
\providecommand \BibitemOpen [0]{}%
\providecommand \bibitemStop [0]{}%
\providecommand \bibitemNoStop [0]{.\EOS\space}%
\providecommand \EOS [0]{\spacefactor3000\relax}%
\providecommand \BibitemShut  [1]{\csname bibitem#1\endcsname}%
\let\auto@bib@innerbib\@empty
\bibitem [{\citenamefont {Benseny}\ \emph {et~al.}(2014)\citenamefont
  {Benseny}, \citenamefont {Albareda}, \citenamefont {Sanz}, \citenamefont
  {Mompart},\ and\ \citenamefont {Oriols}}]{benseny}%
  \BibitemOpen
  \bibfield  {author} {\bibinfo {author} {\bibfnamefont {A.}~\bibnamefont
  {Benseny}}, \bibinfo {author} {\bibfnamefont {G.}~\bibnamefont {Albareda}},
  \bibinfo {author} {\bibfnamefont {{\'A}.~S.}\ \bibnamefont {Sanz}}, \bibinfo
  {author} {\bibfnamefont {J.}~\bibnamefont {Mompart}}, \ and\ \bibinfo
  {author} {\bibfnamefont {X.}~\bibnamefont {Oriols}},\ }\href@noop {}
  {\bibfield  {journal} {\bibinfo  {journal} {Eur.Phys.J. D}\ }\textbf
  {\bibinfo {volume} {68}},\ \bibinfo {pages} {1} (\bibinfo {year}
  {2014})}\BibitemShut {NoStop}%
\bibitem [{\citenamefont {Hirschfelder}\ \emph {et~al.}(1974)\citenamefont
  {Hirschfelder}, \citenamefont {Christoph},\ and\ \citenamefont
  {Palke}}]{hirschfelder1974quantum}%
  \BibitemOpen
  \bibfield  {author} {\bibinfo {author} {\bibfnamefont {J.~O.}\ \bibnamefont
  {Hirschfelder}}, \bibinfo {author} {\bibfnamefont {A.~C.}\ \bibnamefont
  {Christoph}}, \ and\ \bibinfo {author} {\bibfnamefont {W.~E.}\ \bibnamefont
  {Palke}},\ }\href@noop {} {\bibfield  {journal} {\bibinfo  {journal} {The
  Journal of Chemical Physics}\ }\textbf {\bibinfo {volume} {61}},\ \bibinfo
  {pages} {5435} (\bibinfo {year} {1974})}\BibitemShut {NoStop}%
\bibitem [{\citenamefont {Dirac}(1931)}]{dirac1931quantised}%
  \BibitemOpen
  \bibfield  {author} {\bibinfo {author} {\bibfnamefont {P.~A.}\ \bibnamefont
  {Dirac}},\ }\href@noop {} {\bibfield  {journal} {\bibinfo  {journal} {Proc.
  Royal Soc. A}\ }\textbf {\bibinfo {volume} {133}},\ \bibinfo {pages} {60}
  (\bibinfo {year} {1931})}\BibitemShut {NoStop}%
\bibitem [{\citenamefont {Fetter}\ and\ \citenamefont
  {Svidzinsky}(2001)}]{fetter2001vortices}%
  \BibitemOpen
  \bibfield  {author} {\bibinfo {author} {\bibfnamefont {A.~L.}\ \bibnamefont
  {Fetter}}\ and\ \bibinfo {author} {\bibfnamefont {A.~A.}\ \bibnamefont
  {Svidzinsky}},\ }\href@noop {} {\bibfield  {journal} {\bibinfo  {journal} {J.
  Phys. Condens. Matter}\ }\textbf {\bibinfo {volume} {13}},\ \bibinfo {pages}
  {R135} (\bibinfo {year} {2001})}\BibitemShut {NoStop}%
\bibitem [{\citenamefont {Sanz}\ and\ \citenamefont
  {Borondo}(2007)}]{sanz2007quantum}%
  \BibitemOpen
  \bibfield  {author} {\bibinfo {author} {\bibfnamefont {A.~S.}\ \bibnamefont
  {Sanz}}\ and\ \bibinfo {author} {\bibfnamefont {F.}~\bibnamefont {Borondo}},\
  }\href@noop {} {\bibfield  {journal} {\bibinfo  {journal} {Eur. Phys. J. D}\
  }\textbf {\bibinfo {volume} {44}},\ \bibinfo {pages} {319} (\bibinfo {year}
  {2007})}\BibitemShut {NoStop}%
\bibitem [{\citenamefont {Babyuk}\ and\ \citenamefont
  {Nechiporuk}(2011)}]{babyuk2011study}%
  \BibitemOpen
  \bibfield  {author} {\bibinfo {author} {\bibfnamefont {D.}~\bibnamefont
  {Babyuk}}\ and\ \bibinfo {author} {\bibfnamefont {V.}~\bibnamefont
  {Nechiporuk}},\ }\href@noop {} {\bibfield  {journal} {\bibinfo  {journal}
  {Rus. J. Phys. Chem. B}\ }\textbf {\bibinfo {volume} {5}},\ \bibinfo {pages}
  {730} (\bibinfo {year} {2011})}\BibitemShut {NoStop}%
\bibitem [{\citenamefont {Chou}\ \emph {et~al.}(2009)\citenamefont {Chou},
  \citenamefont {Sanz}, \citenamefont {Miret-Art{\'e}s},\ and\ \citenamefont
  {Wyatt}}]{chou2009hydrodynamic}%
  \BibitemOpen
  \bibfield  {author} {\bibinfo {author} {\bibfnamefont {C.-C.}\ \bibnamefont
  {Chou}}, \bibinfo {author} {\bibfnamefont {{\'A}.~S.}\ \bibnamefont {Sanz}},
  \bibinfo {author} {\bibfnamefont {S.}~\bibnamefont {Miret-Art{\'e}s}}, \ and\
  \bibinfo {author} {\bibfnamefont {R.~E.}\ \bibnamefont {Wyatt}},\ }\href@noop
  {} {\bibfield  {journal} {\bibinfo  {journal} {Phys. Rev. Lett.}\ }\textbf
  {\bibinfo {volume} {102}},\ \bibinfo {pages} {250401} (\bibinfo {year}
  {2009})}\BibitemShut {NoStop}%
\bibitem [{\citenamefont {Bell}(2013)}]{bell2013tunnel}%
  \BibitemOpen
  \bibfield  {author} {\bibinfo {author} {\bibfnamefont {R.~P.}\ \bibnamefont
  {Bell}},\ }\href@noop {} {\emph {\bibinfo {title} {The tunnel effect in
  chemistry}}}\ (\bibinfo  {publisher} {Springer},\ \bibinfo {year}
  {2013})\BibitemShut {NoStop}%
\bibitem [{\citenamefont {Curchod}\ \emph {et~al.}(2011)\citenamefont
  {Curchod}, \citenamefont {Tavernelli},\ and\ \citenamefont
  {Rothlisberger}}]{curchod2011trajectory}%
  \BibitemOpen
  \bibfield  {author} {\bibinfo {author} {\bibfnamefont {B.~F.}\ \bibnamefont
  {Curchod}}, \bibinfo {author} {\bibfnamefont {I.}~\bibnamefont {Tavernelli}},
  \ and\ \bibinfo {author} {\bibfnamefont {U.}~\bibnamefont {Rothlisberger}},\
  }\href@noop {} {\bibfield  {journal} {\bibinfo  {journal} {Phys. Chem. Chem.
  Phys.}\ }\textbf {\bibinfo {volume} {13}},\ \bibinfo {pages} {3231} (\bibinfo
  {year} {2011})}\BibitemShut {NoStop}%
\bibitem [{\citenamefont {Garashchuk}\ \emph {et~al.}(2015)\citenamefont
  {Garashchuk}, \citenamefont {Jakowski},\ and\ \citenamefont
  {Rassolov}}]{garashchuk2015approximate}%
  \BibitemOpen
  \bibfield  {author} {\bibinfo {author} {\bibfnamefont {S.}~\bibnamefont
  {Garashchuk}}, \bibinfo {author} {\bibfnamefont {J.}~\bibnamefont
  {Jakowski}}, \ and\ \bibinfo {author} {\bibfnamefont {V.~A.}\ \bibnamefont
  {Rassolov}},\ }\href@noop {} {\bibfield  {journal} {\bibinfo  {journal} {Mol.
  Simul.}\ }\textbf {\bibinfo {volume} {41}},\ \bibinfo {pages} {86} (\bibinfo
  {year} {2015})}\BibitemShut {NoStop}%
\bibitem [{\citenamefont {Rudinsky}\ \emph {et~al.}(2017)\citenamefont
  {Rudinsky}, \citenamefont {Sanz},\ and\ \citenamefont
  {Gauvin}}]{rudinsky2017novel}%
  \BibitemOpen
  \bibfield  {author} {\bibinfo {author} {\bibfnamefont {S.}~\bibnamefont
  {Rudinsky}}, \bibinfo {author} {\bibfnamefont {A.~S.}\ \bibnamefont {Sanz}},
  \ and\ \bibinfo {author} {\bibfnamefont {R.}~\bibnamefont {Gauvin}},\
  }\href@noop {} {\bibfield  {journal} {\bibinfo  {journal} {J. Chem. Phys.}\
  }\textbf {\bibinfo {volume} {146}},\ \bibinfo {pages} {104702} (\bibinfo
  {year} {2017})}\BibitemShut {NoStop}%
\bibitem [{\citenamefont {Atzmon}\ and\ \citenamefont
  {Shimshoni}(2012)}]{atzmon2012alternating}%
  \BibitemOpen
  \bibfield  {author} {\bibinfo {author} {\bibfnamefont {Y.}~\bibnamefont
  {Atzmon}}\ and\ \bibinfo {author} {\bibfnamefont {E.}~\bibnamefont
  {Shimshoni}},\ }\href@noop {} {\bibfield  {journal} {\bibinfo  {journal}
  {Phys. Rev. B}\ }\textbf {\bibinfo {volume} {85}},\ \bibinfo {pages} {134523}
  (\bibinfo {year} {2012})}\BibitemShut {NoStop}%
\bibitem [{\citenamefont {Fazio}\ and\ \citenamefont
  {Sch{\"o}n}(2013)}]{fazio2013charges}%
  \BibitemOpen
  \bibfield  {author} {\bibinfo {author} {\bibfnamefont {R.}~\bibnamefont
  {Fazio}}\ and\ \bibinfo {author} {\bibfnamefont {G.}~\bibnamefont
  {Sch{\"o}n}},\ }in\ \href@noop {} {\emph {\bibinfo {booktitle} {40 Years of
  Berezinskii--Kosterlitz--Thouless Theory}}}\ (\bibinfo  {publisher} {World
  Scientific},\ \bibinfo {year} {2013})\ pp.\ \bibinfo {pages}
  {237--254}\BibitemShut {NoStop}%
\bibitem [{\citenamefont {Madelung}(1927)}]{madelung1927quantentheorie}%
  \BibitemOpen
  \bibfield  {author} {\bibinfo {author} {\bibfnamefont {E.}~\bibnamefont
  {Madelung}},\ }\href@noop {} {\bibfield  {journal} {\bibinfo  {journal}
  {Zeit. Phys. A}\ }\textbf {\bibinfo {volume} {40}},\ \bibinfo {pages} {322}
  (\bibinfo {year} {1927})}\BibitemShut {NoStop}%
\bibitem [{\citenamefont {Bohm}(1952{\natexlab{a}})}]{Bohm}%
  \BibitemOpen
  \bibfield  {author} {\bibinfo {author} {\bibfnamefont {D.}~\bibnamefont
  {Bohm}},\ }\href@noop {} {\bibfield  {journal} {\bibinfo  {journal} {Phys.
  Rev.}\ }\textbf {\bibinfo {volume} {85}},\ \bibinfo {pages} {166} (\bibinfo
  {year} {1952}{\natexlab{a}})}\BibitemShut {NoStop}%
\bibitem [{\citenamefont {Bohm}(1952{\natexlab{b}})}]{BohmII}%
  \BibitemOpen
  \bibfield  {author} {\bibinfo {author} {\bibfnamefont {D.}~\bibnamefont
  {Bohm}},\ }\href@noop {} {\bibfield  {journal} {\bibinfo  {journal} {Phys.
  Rev.}\ }\textbf {\bibinfo {volume} {85}},\ \bibinfo {pages} {180} (\bibinfo
  {year} {1952}{\natexlab{b}})}\BibitemShut {NoStop}%
\bibitem [{\citenamefont {Trahan}\ and\ \citenamefont
  {Wyatt}(2005)}]{trahan2005quantum}%
  \BibitemOpen
  \bibfield  {author} {\bibinfo {author} {\bibfnamefont {C.}~\bibnamefont
  {Trahan}}\ and\ \bibinfo {author} {\bibfnamefont {R.}~\bibnamefont {Wyatt}},\
  }\href@noop {} {\emph {\bibinfo {title} {Quantum Dynamics with Trajectories:
  Introduction to Quantum Hydrodynamics,}}},\ Interdisciplinary Applied
  Mathematics\ (\bibinfo  {publisher} {Springer},\ \bibinfo {year}
  {2005})\BibitemShut {NoStop}%
\bibitem [{\citenamefont {Sanz}\ and\ \citenamefont
  {Miret-Art{\'e}s}(2012)}]{sanz2012trajectory}%
  \BibitemOpen
  \bibfield  {author} {\bibinfo {author} {\bibfnamefont {{\'A}.~S.}\
  \bibnamefont {Sanz}}\ and\ \bibinfo {author} {\bibfnamefont {S.}~\bibnamefont
  {Miret-Art{\'e}s}},\ }\href@noop {} {\emph {\bibinfo {title} {A Trajectory
  Description of Quantum Processes. I. Fundamentals: A {B}ohmian
  Perspective}}},\ LNP\ (\bibinfo  {publisher} {Springer},\ \bibinfo {year}
  {2012})\BibitemShut {NoStop}%
\bibitem [{\citenamefont {Sanz}\ and\ \citenamefont
  {Miret-Art{\'e}s}(2013)}]{sanz2013trajectory}%
  \BibitemOpen
  \bibfield  {author} {\bibinfo {author} {\bibfnamefont {{\'A}.}~\bibnamefont
  {Sanz}}\ and\ \bibinfo {author} {\bibfnamefont {S.}~\bibnamefont
  {Miret-Art{\'e}s}},\ }\href@noop {} {\emph {\bibinfo {title} {A Trajectory
  Description of Quantum Processes. II. Applications: A Bohmian
  Perspective}}},\ LNP\ (\bibinfo  {publisher} {Springer},\ \bibinfo {year}
  {2013})\BibitemShut {NoStop}%
\bibitem [{\citenamefont {Deotto}\ and\ \citenamefont
  {Ghirardi}(1998)}]{deotto1998bohmian}%
  \BibitemOpen
  \bibfield  {author} {\bibinfo {author} {\bibfnamefont {E.}~\bibnamefont
  {Deotto}}\ and\ \bibinfo {author} {\bibfnamefont {G.~C.}\ \bibnamefont
  {Ghirardi}},\ }\href@noop {} {\bibfield  {journal} {\bibinfo  {journal}
  {Found. Phys.}\ }\textbf {\bibinfo {volume} {28}},\ \bibinfo {pages} {1}
  (\bibinfo {year} {1998})}\BibitemShut {NoStop}%
\bibitem [{\citenamefont {Vaidman}(2005)}]{vaidman2005reality}%
  \BibitemOpen
  \bibfield  {author} {\bibinfo {author} {\bibfnamefont {L.}~\bibnamefont
  {Vaidman}},\ }\href@noop {} {\bibfield  {journal} {\bibinfo  {journal}
  {Found. Phys.}\ }\textbf {\bibinfo {volume} {35}},\ \bibinfo {pages} {299}
  (\bibinfo {year} {2005})}\BibitemShut {NoStop}%
\bibitem [{\citenamefont {D{\"u}rr}\ and\ \citenamefont
  {Teufel}(2009)}]{durr2009bohmian}%
  \BibitemOpen
  \bibfield  {author} {\bibinfo {author} {\bibfnamefont {D.}~\bibnamefont
  {D{\"u}rr}}\ and\ \bibinfo {author} {\bibfnamefont {S.}~\bibnamefont
  {Teufel}},\ }\href@noop {} {\emph {\bibinfo {title} {Bohmian mechanics: The
  Physics and Mathematics of Quantum Theory}}}\ (\bibinfo  {publisher}
  {Springer},\ \bibinfo {year} {2009})\BibitemShut {NoStop}%
\bibitem [{\citenamefont {Kocsis}\ \emph {et~al.}(2011)\citenamefont {Kocsis},
  \citenamefont {Braverman}, \citenamefont {Ravets}, \citenamefont {Stevens},
  \citenamefont {Mirin}, \citenamefont {Shalm},\ and\ \citenamefont
  {Steinberg}}]{kocsis2011observing}%
  \BibitemOpen
  \bibfield  {author} {\bibinfo {author} {\bibfnamefont {S.}~\bibnamefont
  {Kocsis}}, \bibinfo {author} {\bibfnamefont {B.}~\bibnamefont {Braverman}},
  \bibinfo {author} {\bibfnamefont {S.}~\bibnamefont {Ravets}}, \bibinfo
  {author} {\bibfnamefont {M.~J.}\ \bibnamefont {Stevens}}, \bibinfo {author}
  {\bibfnamefont {R.~P.}\ \bibnamefont {Mirin}}, \bibinfo {author}
  {\bibfnamefont {L.~K.}\ \bibnamefont {Shalm}}, \ and\ \bibinfo {author}
  {\bibfnamefont {A.~M.}\ \bibnamefont {Steinberg}},\ }\href@noop {} {\bibfield
   {journal} {\bibinfo  {journal} {Science}\ }\textbf {\bibinfo {volume}
  {332}},\ \bibinfo {pages} {1170} (\bibinfo {year} {2011})}\BibitemShut
  {NoStop}%
\bibitem [{\citenamefont {Mahler}\ \emph {et~al.}(2016)\citenamefont {Mahler},
  \citenamefont {Rozema}, \citenamefont {Fisher}, \citenamefont {Vermeyden},
  \citenamefont {Resch}, \citenamefont {Wiseman},\ and\ \citenamefont
  {Steinberg}}]{mahler2016experimental}%
  \BibitemOpen
  \bibfield  {author} {\bibinfo {author} {\bibfnamefont {D.~H.}\ \bibnamefont
  {Mahler}}, \bibinfo {author} {\bibfnamefont {L.}~\bibnamefont {Rozema}},
  \bibinfo {author} {\bibfnamefont {K.}~\bibnamefont {Fisher}}, \bibinfo
  {author} {\bibfnamefont {L.}~\bibnamefont {Vermeyden}}, \bibinfo {author}
  {\bibfnamefont {K.~J.}\ \bibnamefont {Resch}}, \bibinfo {author}
  {\bibfnamefont {H.~M.}\ \bibnamefont {Wiseman}}, \ and\ \bibinfo {author}
  {\bibfnamefont {A.}~\bibnamefont {Steinberg}},\ }\href@noop {} {\bibfield
  {journal} {\bibinfo  {journal} {Sci. Adv.}\ }\textbf {\bibinfo {volume}
  {2}},\ \bibinfo {pages} {e1501466} (\bibinfo {year} {2016})}\BibitemShut
  {NoStop}%
\bibitem [{\citenamefont {Gisin}(2018)}]{gisin2018bohmian}%
  \BibitemOpen
  \bibfield  {author} {\bibinfo {author} {\bibfnamefont {N.}~\bibnamefont
  {Gisin}},\ }\href@noop {} {\bibfield  {journal} {\bibinfo  {journal}
  {Entropy}\ }\textbf {\bibinfo {volume} {20}},\ \bibinfo {pages} {105}
  (\bibinfo {year} {2018})}\BibitemShut {NoStop}%
\bibitem [{\citenamefont {Wu}\ and\ \citenamefont
  {Sprung}(1994)}]{wu1994inverse}%
  \BibitemOpen
  \bibfield  {author} {\bibinfo {author} {\bibfnamefont {H.}~\bibnamefont
  {Wu}}\ and\ \bibinfo {author} {\bibfnamefont {D.}~\bibnamefont {Sprung}},\
  }\href@noop {} {\bibfield  {journal} {\bibinfo  {journal} {Phys. Rev. A}\
  }\textbf {\bibinfo {volume} {49}},\ \bibinfo {pages} {4305} (\bibinfo {year}
  {1994})}\BibitemShut {NoStop}%
\bibitem [{\citenamefont {Frisk}(1997)}]{frisk1997properties}%
  \BibitemOpen
  \bibfield  {author} {\bibinfo {author} {\bibfnamefont {H.}~\bibnamefont
  {Frisk}},\ }\href@noop {} {\bibfield  {journal} {\bibinfo  {journal} {Phys.
  Lett. A}\ }\textbf {\bibinfo {volume} {227}},\ \bibinfo {pages} {139}
  (\bibinfo {year} {1997})}\BibitemShut {NoStop}%
\bibitem [{\citenamefont {Wu}\ and\ \citenamefont
  {Sprung}(1999)}]{wu1999quantum}%
  \BibitemOpen
  \bibfield  {author} {\bibinfo {author} {\bibfnamefont {H.}~\bibnamefont
  {Wu}}\ and\ \bibinfo {author} {\bibfnamefont {D.}~\bibnamefont {Sprung}},\
  }\href@noop {} {\bibfield  {journal} {\bibinfo  {journal} {Phys. Lett. A}\
  }\textbf {\bibinfo {volume} {261}},\ \bibinfo {pages} {150} (\bibinfo {year}
  {1999})}\BibitemShut {NoStop}%
\bibitem [{\citenamefont {Falsaperla}\ and\ \citenamefont
  {Fonte}(2003)}]{falsaperla2003motion}%
  \BibitemOpen
  \bibfield  {author} {\bibinfo {author} {\bibfnamefont {P.}~\bibnamefont
  {Falsaperla}}\ and\ \bibinfo {author} {\bibfnamefont {G.}~\bibnamefont
  {Fonte}},\ }\href@noop {} {\bibfield  {journal} {\bibinfo  {journal} {Phys.
  Lett. A}\ }\textbf {\bibinfo {volume} {316}},\ \bibinfo {pages} {382}
  (\bibinfo {year} {2003})}\BibitemShut {NoStop}%
\bibitem [{\citenamefont {Wisniacki}\ and\ \citenamefont
  {Pujals}(2005)}]{wisniacki2005motion}%
  \BibitemOpen
  \bibfield  {author} {\bibinfo {author} {\bibfnamefont {D.~A.}\ \bibnamefont
  {Wisniacki}}\ and\ \bibinfo {author} {\bibfnamefont {E.~R.}\ \bibnamefont
  {Pujals}},\ }\href@noop {} {\bibfield  {journal} {\bibinfo  {journal}
  {Europhys. Lett.}\ }\textbf {\bibinfo {volume} {71}},\ \bibinfo {pages} {159}
  (\bibinfo {year} {2005})}\BibitemShut {NoStop}%
\bibitem [{\citenamefont {Wisniacki}\ \emph {et~al.}(2007)\citenamefont
  {Wisniacki}, \citenamefont {Pujals},\ and\ \citenamefont
  {Borondo}}]{wisniacki2007vortex}%
  \BibitemOpen
  \bibfield  {author} {\bibinfo {author} {\bibfnamefont {D.}~\bibnamefont
  {Wisniacki}}, \bibinfo {author} {\bibfnamefont {E.}~\bibnamefont {Pujals}}, \
  and\ \bibinfo {author} {\bibfnamefont {F.}~\bibnamefont {Borondo}},\
  }\href@noop {} {\bibfield  {journal} {\bibinfo  {journal} {J. Phys. A.}\
  }\textbf {\bibinfo {volume} {40}},\ \bibinfo {pages} {14353} (\bibinfo {year}
  {2007})}\BibitemShut {NoStop}%
\bibitem [{\citenamefont {Efthymiopoulos}\ \emph {et~al.}(2007)\citenamefont
  {Efthymiopoulos}, \citenamefont {Kalapotharakos},\ and\ \citenamefont
  {Contopoulos}}]{efthymiopoulos2007nodal}%
  \BibitemOpen
  \bibfield  {author} {\bibinfo {author} {\bibfnamefont {C.}~\bibnamefont
  {Efthymiopoulos}}, \bibinfo {author} {\bibfnamefont {C.}~\bibnamefont
  {Kalapotharakos}}, \ and\ \bibinfo {author} {\bibfnamefont {G.}~\bibnamefont
  {Contopoulos}},\ }\href@noop {} {\bibfield  {journal} {\bibinfo  {journal}
  {J. Phys. A}\ }\textbf {\bibinfo {volume} {40}},\ \bibinfo {pages} {12945}
  (\bibinfo {year} {2007})}\BibitemShut {NoStop}%
\bibitem [{\citenamefont {Efthymiopoulos}\ \emph {et~al.}(2009)\citenamefont
  {Efthymiopoulos}, \citenamefont {Kalapotharakos},\ and\ \citenamefont
  {Contopoulos}}]{Efth2009}%
  \BibitemOpen
  \bibfield  {author} {\bibinfo {author} {\bibfnamefont {C.}~\bibnamefont
  {Efthymiopoulos}}, \bibinfo {author} {\bibfnamefont {C.}~\bibnamefont
  {Kalapotharakos}}, \ and\ \bibinfo {author} {\bibfnamefont {G.}~\bibnamefont
  {Contopoulos}},\ }\href@noop {} {\bibfield  {journal} {\bibinfo  {journal}
  {Phys. Rev. E}\ }\textbf {\bibinfo {volume} {79}},\ \bibinfo {pages} {036203}
  (\bibinfo {year} {2009})}\BibitemShut {NoStop}%
\bibitem [{\citenamefont {Cesa}\ \emph {et~al.}(2016)\citenamefont {Cesa},
  \citenamefont {Martin},\ and\ \citenamefont {Struyve}}]{cesa2016chaotic}%
  \BibitemOpen
  \bibfield  {author} {\bibinfo {author} {\bibfnamefont {A.}~\bibnamefont
  {Cesa}}, \bibinfo {author} {\bibfnamefont {J.}~\bibnamefont {Martin}}, \ and\
  \bibinfo {author} {\bibfnamefont {W.}~\bibnamefont {Struyve}},\ }\href@noop
  {} {\bibfield  {journal} {\bibinfo  {journal} {J. Phys. A}\ }\textbf
  {\bibinfo {volume} {49}},\ \bibinfo {pages} {395301} (\bibinfo {year}
  {2016})}\BibitemShut {NoStop}%
\bibitem [{\citenamefont {Lopreore}\ and\ \citenamefont
  {Wyatt}(1999)}]{lopreore1999quantum}%
  \BibitemOpen
  \bibfield  {author} {\bibinfo {author} {\bibfnamefont {C.~L.}\ \bibnamefont
  {Lopreore}}\ and\ \bibinfo {author} {\bibfnamefont {R.~E.}\ \bibnamefont
  {Wyatt}},\ }\href@noop {} {\bibfield  {journal} {\bibinfo  {journal} {Phys.
  Rev. Lett.}\ }\textbf {\bibinfo {volume} {82}},\ \bibinfo {pages} {5190}
  (\bibinfo {year} {1999})}\BibitemShut {NoStop}%
\bibitem [{\citenamefont {Babyuk}\ \emph {et~al.}(2003)\citenamefont {Babyuk},
  \citenamefont {Wyatt},\ and\ \citenamefont
  {Frederick}}]{babyuk2003hydrodynamic}%
  \BibitemOpen
  \bibfield  {author} {\bibinfo {author} {\bibfnamefont {D.}~\bibnamefont
  {Babyuk}}, \bibinfo {author} {\bibfnamefont {R.~E.}\ \bibnamefont {Wyatt}}, \
  and\ \bibinfo {author} {\bibfnamefont {J.~H.}\ \bibnamefont {Frederick}},\
  }\href@noop {} {\bibfield  {journal} {\bibinfo  {journal} {J. Chem. Phys.}\
  }\textbf {\bibinfo {volume} {119}},\ \bibinfo {pages} {6482} (\bibinfo {year}
  {2003})}\BibitemShut {NoStop}%
\bibitem [{\citenamefont {Alarc{\'o}n}\ \emph {et~al.}(2013)\citenamefont
  {Alarc{\'o}n}, \citenamefont {Yaro}, \citenamefont {Cartoixa},\ and\
  \citenamefont {Oriols}}]{alarcon2013computation}%
  \BibitemOpen
  \bibfield  {author} {\bibinfo {author} {\bibfnamefont {A.}~\bibnamefont
  {Alarc{\'o}n}}, \bibinfo {author} {\bibfnamefont {S.}~\bibnamefont {Yaro}},
  \bibinfo {author} {\bibfnamefont {X.}~\bibnamefont {Cartoixa}}, \ and\
  \bibinfo {author} {\bibfnamefont {X.}~\bibnamefont {Oriols}},\ }\href@noop {}
  {\bibfield  {journal} {\bibinfo  {journal} {J. Phys. Condens. Matter}\
  }\textbf {\bibinfo {volume} {25}},\ \bibinfo {pages} {325601} (\bibinfo
  {year} {2013})}\BibitemShut {NoStop}%
\bibitem [{\citenamefont {Marian}\ \emph {et~al.}(2016)\citenamefont {Marian},
  \citenamefont {Oriols},\ and\ \citenamefont {Zangh{\`\i}}}]{marian2016noise}%
  \BibitemOpen
  \bibfield  {author} {\bibinfo {author} {\bibfnamefont {D.}~\bibnamefont
  {Marian}}, \bibinfo {author} {\bibfnamefont {X.}~\bibnamefont {Oriols}}, \
  and\ \bibinfo {author} {\bibfnamefont {N.}~\bibnamefont {Zangh{\`\i}}},\
  }\href@noop {} {\bibfield  {journal} {\bibinfo  {journal} {J. Stat. Mech.
  Theory Exp.}\ }\textbf {\bibinfo {volume} {2016}},\ \bibinfo {pages} {054011}
  (\bibinfo {year} {2016})}\BibitemShut {NoStop}%
\bibitem [{\citenamefont {Colom{\'e}s}\ \emph {et~al.}(2017)\citenamefont
  {Colom{\'e}s}, \citenamefont {Zhan}, \citenamefont {Marian},\ and\
  \citenamefont {Oriols}}]{colomes2017quantum}%
  \BibitemOpen
  \bibfield  {author} {\bibinfo {author} {\bibfnamefont {E.}~\bibnamefont
  {Colom{\'e}s}}, \bibinfo {author} {\bibfnamefont {Z.}~\bibnamefont {Zhan}},
  \bibinfo {author} {\bibfnamefont {D.}~\bibnamefont {Marian}}, \ and\ \bibinfo
  {author} {\bibfnamefont {X.}~\bibnamefont {Oriols}},\ }\href@noop {}
  {\bibfield  {journal} {\bibinfo  {journal} {Phys. Rev. B}\ }\textbf {\bibinfo
  {volume} {96}},\ \bibinfo {pages} {075135} (\bibinfo {year}
  {2017})}\BibitemShut {NoStop}%
\bibitem [{\citenamefont {Efthymiopoulos}\ \emph {et~al.}(2017)\citenamefont
  {Efthymiopoulos}, \citenamefont {Contopoulos},\ and\ \citenamefont
  {Tzemos}}]{efthymiopoulos2017chaos}%
  \BibitemOpen
  \bibfield  {author} {\bibinfo {author} {\bibfnamefont {C.}~\bibnamefont
  {Efthymiopoulos}}, \bibinfo {author} {\bibfnamefont {G.}~\bibnamefont
  {Contopoulos}}, \ and\ \bibinfo {author} {\bibfnamefont {A.~C.}\ \bibnamefont
  {Tzemos}},\ }\href@noop {} {\bibfield  {journal} {\bibinfo  {journal} {Ann.
  Fond. Louis de Broglie}\ }\textbf {\bibinfo {volume} {42}},\ \bibinfo {pages}
  {133} (\bibinfo {year} {2017})}\BibitemShut {NoStop}%
\bibitem [{\citenamefont {Valentini}\ and\ \citenamefont
  {Westman}(2005)}]{valentini2005dynamical}%
  \BibitemOpen
  \bibfield  {author} {\bibinfo {author} {\bibfnamefont {A.}~\bibnamefont
  {Valentini}}\ and\ \bibinfo {author} {\bibfnamefont {H.}~\bibnamefont
  {Westman}},\ }\href@noop {} {\bibfield  {journal} {\bibinfo  {journal} {Proc.
  Royal Soc. A}\ }\textbf {\bibinfo {volume} {461}},\ \bibinfo {pages} {253}
  (\bibinfo {year} {2005})}\BibitemShut {NoStop}%
\bibitem [{\citenamefont {Efthymiopoulos}\ and\ \citenamefont
  {Contopoulos}(2006)}]{efthymiopoulos2006chaos}%
  \BibitemOpen
  \bibfield  {author} {\bibinfo {author} {\bibfnamefont {C.}~\bibnamefont
  {Efthymiopoulos}}\ and\ \bibinfo {author} {\bibfnamefont {G.}~\bibnamefont
  {Contopoulos}},\ }\href@noop {} {\bibfield  {journal} {\bibinfo  {journal}
  {J. Phys. A}\ }\textbf {\bibinfo {volume} {39}},\ \bibinfo {pages} {1819}
  (\bibinfo {year} {2006})}\BibitemShut {NoStop}%
\bibitem [{\citenamefont {D{\"u}rr}\ \emph {et~al.}(1992)\citenamefont
  {D{\"u}rr}, \citenamefont {Goldstein},\ and\ \citenamefont
  {Zanghi}}]{durr1992quantum}%
  \BibitemOpen
  \bibfield  {author} {\bibinfo {author} {\bibfnamefont {D.}~\bibnamefont
  {D{\"u}rr}}, \bibinfo {author} {\bibfnamefont {S.}~\bibnamefont {Goldstein}},
  \ and\ \bibinfo {author} {\bibfnamefont {N.}~\bibnamefont {Zanghi}},\
  }\href@noop {} {\bibfield  {journal} {\bibinfo  {journal} {J. Stat. Phys.}\
  }\textbf {\bibinfo {volume} {68}},\ \bibinfo {pages} {259} (\bibinfo {year}
  {1992})}\BibitemShut {NoStop}%
\bibitem [{\citenamefont {Borondo}\ \emph {et~al.}(2009)\citenamefont
  {Borondo}, \citenamefont {Luque}, \citenamefont {Villanueva},\ and\
  \citenamefont {Wisniacki}}]{borondo2009dynamical}%
  \BibitemOpen
  \bibfield  {author} {\bibinfo {author} {\bibfnamefont {F.}~\bibnamefont
  {Borondo}}, \bibinfo {author} {\bibfnamefont {A.}~\bibnamefont {Luque}},
  \bibinfo {author} {\bibfnamefont {J.}~\bibnamefont {Villanueva}}, \ and\
  \bibinfo {author} {\bibfnamefont {D.~A.}\ \bibnamefont {Wisniacki}},\
  }\href@noop {} {\bibfield  {journal} {\bibinfo  {journal} {J. Phys. A}\
  }\textbf {\bibinfo {volume} {42}},\ \bibinfo {pages} {495103} (\bibinfo
  {year} {2009})}\BibitemShut {NoStop}%
\bibitem [{\citenamefont {Tzemos}\ \emph {et~al.}(2016)\citenamefont {Tzemos},
  \citenamefont {Contopoulos},\ and\ \citenamefont
  {Efthymiopoulos}}]{Tzemos2016}%
  \BibitemOpen
  \bibfield  {author} {\bibinfo {author} {\bibfnamefont {A.~C.}\ \bibnamefont
  {Tzemos}}, \bibinfo {author} {\bibfnamefont {G.}~\bibnamefont {Contopoulos}},
  \ and\ \bibinfo {author} {\bibfnamefont {C.}~\bibnamefont {Efthymiopoulos}},\
  }\href@noop {} {\bibfield  {journal} {\bibinfo  {journal} {Phys. Lett. A}\
  }\textbf {\bibinfo {volume} {380}},\ \bibinfo {pages} {3796} (\bibinfo {year}
  {2016})}\BibitemShut {NoStop}%
\bibitem [{\citenamefont {Contopoulos}\ \emph {et~al.}(2017)\citenamefont
  {Contopoulos}, \citenamefont {Tzemos},\ and\ \citenamefont
  {Efthymiopoulos}}]{contopoulos2017partial}%
  \BibitemOpen
  \bibfield  {author} {\bibinfo {author} {\bibfnamefont {G.}~\bibnamefont
  {Contopoulos}}, \bibinfo {author} {\bibfnamefont {A.~C.}\ \bibnamefont
  {Tzemos}}, \ and\ \bibinfo {author} {\bibfnamefont {C.}~\bibnamefont
  {Efthymiopoulos}},\ }\href@noop {} {\bibfield  {journal} {\bibinfo  {journal}
  {J. Phys. A}\ }\textbf {\bibinfo {volume} {50}},\ \bibinfo {pages} {195101}
  (\bibinfo {year} {2017})}\BibitemShut {NoStop}%
\bibitem [{\citenamefont {Wiggins}(2013)}]{wiggins2013normally}%
  \BibitemOpen
  \bibfield  {author} {\bibinfo {author} {\bibfnamefont {S.}~\bibnamefont
  {Wiggins}},\ }\href@noop {} {\emph {\bibinfo {title} {Normally hyperbolic
  invariant manifolds in dynamical systems}}},\ Vol.\ \bibinfo {volume} {105}\
  (\bibinfo  {publisher} {Springer},\ \bibinfo {year} {2013})\BibitemShut
  {NoStop}%
\bibitem [{\citenamefont {Voglis}\ and\ \citenamefont
  {Contopoulos}(1994)}]{voglis1994invariant}%
  \BibitemOpen
  \bibfield  {author} {\bibinfo {author} {\bibfnamefont {N.}~\bibnamefont
  {Voglis}}\ and\ \bibinfo {author} {\bibfnamefont {G.}~\bibnamefont
  {Contopoulos}},\ }\href@noop {} {\bibfield  {journal} {\bibinfo  {journal}
  {J. Phys. A}\ }\textbf {\bibinfo {volume} {27}},\ \bibinfo {pages} {4899}
  (\bibinfo {year} {1994})}\BibitemShut {NoStop}%
\bibitem [{\citenamefont {Contopoulos}(2002)}]{Contopoulos200210}%
  \BibitemOpen
  \bibfield  {author} {\bibinfo {author} {\bibfnamefont {G.}~\bibnamefont
  {Contopoulos}},\ }\href@noop {} {\emph {\bibinfo {title} {Order and Chaos in
  Dynamical Astronomy}}}\ (\bibinfo  {publisher} {Springer},\ \bibinfo {year}
  {2002})\BibitemShut {NoStop}%
\bibitem [{\citenamefont {Strogatz}(2014)}]{strogatz2014nonlinear}%
  \BibitemOpen
  \bibfield  {author} {\bibinfo {author} {\bibfnamefont {S.~H.}\ \bibnamefont
  {Strogatz}},\ }\href@noop {} {\emph {\bibinfo {title} {Nonlinear dynamics and
  chaos: with applications to physics, biology, chemistry, and engineering}}}\
  (\bibinfo  {publisher} {Hachette UK},\ \bibinfo {year} {2014})\BibitemShut
  {NoStop}%
\end{thebibliography}%


%

\end{document}